\documentclass[10pt,twocolumn]{article}

\usepackage[dvipdfmx]{graphicx} %XXX: for only platex?
\usepackage[hyphens]{url}
\usepackage{amsfonts} % for \checkmark

\usepackage{array, ragged2e}
\newcolumntype{R}[1]{>{\RaggedLeft\arraybackslash}p{#1}}
\newcolumntype{P}[1]{>{\centering\arraybackslash}p{#1}}

\newcommand\figref[1]{Figure~\ref{#1}}
\newcommand\tabref[1]{Table~\ref{#1}}

\usepackage{listings}
\usepackage{xcolor}
\usepackage{fancybox}
\usepackage{framed}
\usepackage{authblk}

\newenvironment{mybullet}{\begin{list}{$\bullet$}
    {\setlength{\topsep}{0mm}\setlength{\itemsep}{0mm}
      \setlength{\parsep}{0.5mm}
      \setlength{\itemindent}{0mm}\setlength{\partopsep}{0mm}
      \setlength{\labelwidth}{15mm}
      \setlength{\leftmargin}{8mm}}}{\end{list}}
\newenvironment{mybullet2}{\begin{list}{$\bullet$}
    {\setlength{\topsep}{0mm}\setlength{\itemsep}{0mm}
      \setlength{\parsep}{0mm}
      \setlength{\itemindent}{0mm}\setlength{\partopsep}{0mm}
      \setlength{\labelwidth}{0mm}
      \setlength{\leftmargin}{2mm}}}{\end{list}}

\newcommand{\ctext}[1]{\raise0.2ex\hbox{\textcircled{\scriptsize{#1}}}}

\begin{document}

%don't want date printed
\date{}

%make title bold and 14 pt font (Latex default is non-bold, 16 pt)
% \title{\Large \textbf Trojan of Things: Understanding the threats of malicious NFC tags embedded in the things}
\title{\Large\textbf Trojan of Things: Embedding Malicious NFC Tags into Common Objects}

\author[1]{Seita Maruyama\thanks{maruyama@nsl.cs.waseda.ac.jp}}
\author[1]{Satohiro Wakabayashi\thanks{wakabayashi@goto.info.waseda.ac.jp}}
\author[1]{Tatsuya Mori\thanks{mori@nsl.cs.waseda.ac.jp}}
\affil[1]{Department of Computer Science, Waseda University}
% {\textrm Seita Maruyama}\\
% Waseda University
% \and
% {\textrm Satohiro Wakabayashi}\\
% Waseda University
%\and
% {\textrm Tatsuya Mori}\\
% Waseda University
% copy the following lines to add more authors
% \and
% {\rm Name}\\
%Name Institution
% } % end author

\maketitle

% Use the following at camera-ready time to suppress page numbers.
% Comment it out when you first submit the paper for review.
% \thispagestyle{empty} %XXX

% \tableofcontents

\subsection*{Abstract}
% Your Abstract Text Goes Here.  Just a few facts.
% Whet our appetites.

We present a novel proof-of-concept attack named {\em Trojan
  of Things} ({\em ToT}), which aims to attack NFC-enabled mobile
devices such as smartphones. 
The key idea of ToT attacks is to covertly embed maliciously programmed
NFC tags into common objects routinely encountered in daily life such
as banknotes, clothing, or furniture, which are {\em not} considered
as NFC touchpoints.
% The maliciously programmed NFC tags can be used for various bad
% purposes; e.g., opening a malicious URL in a browser without user's
% consensus or forcing a smartphone to connect to a Wi-Fi AP that
% employs man-in-the-middle attack. 
To fully explore the threat of ToT, we develop two striking
techniques named {\em ToT device} and {\em Phantom touch generator}.
These techniques enable an attacker to carry out various severe and
sophisticated attacks unbeknownst to the device owner who
unintentionally puts the device close to a ToT.
We discuss the feasibility of the attack as well as the possible
countermeasures against the threats of ToT attacks.

\section{Introduction}
\label{sec:intro}

% 1. スマートフォンが色々なものにつながってしまう状況にある
% ことを述べる paragraph
Today, we use a smartphone not only for accessing to the various
Internet services, but also for interacting with the networked devices
around us, e.g., wireless headphones, fitness devices, smart home
devices, connected cars, and contactless payment systems.
To communicate with these networked devices, modern smartphones are
shipped with various networking interfaces such as cellular
networks, Wi-Fi, Bluetooth, and NFC. This trend has made
smartphones getting more and more connected to our life  --
Anywhere, Anytime and with Anything. 

% 2. そのような状況で起こりうる脅威として、ToT を提唱。
Given the pervasive network connectivity of smartphones, we propose a
new proof-of-concept attack named {\em Trojan of Things} ({\em ToT}). 
ToT attacks target the NFC-enabled mobile devices such
smartphones. 
The key idea of ToT attacks is to covertly embed maliciously programmed
NFC tags into common objects (``things'') routinely encountered in
daily life such as banknotes, clothing, or furniture, which are 
{\em not} considered as NFC touchpoints.
NFC tags are passive devices that can communicate with active
NFC devices, e.g., NFC-equipped smartphones.
An NFC tag comprises a thin processor chip and an antenna.
It is small enough to be embedded into a business card. 

% 3. ToT によって持たされる基本的な脅威と既存研究との違い
The threat of maliciously programmed NFC tag has been reported in the
past~\cite{rieback2006your,mulliner2009vulnerability,verdult2011practical,miller2012don,wallofsheep,gold2015testbed}.
An attacker can leverage an NFC tag to trigger risky actions; 
e.g., opening a malicious URL in a browser without user
approval~\cite{miller2012don} or forcing a smartphone to pair with a
rogue Bluetooth device~\cite{verdult2011practical, miller2012don}.
% で述べられていますが、NFCタグによる悪性Wi-Fi APへの接続については参考文献が見つかりません。
% employs a man-in-the-middle attack%~\cite{}. 
% Bluetoothのペアリングにより攻撃が可能であることは~\cite{verdult2011practical, miller2012don}で述べられていますが、NFCタグによる悪性Wi-Fi APへの接続については参考文献が見つかりません。
What distinguishes ToT from prior work is its stealth; instead
of actively prompting victims to touch a point such as a smart poster
where a malicious NFC tag was embedded, ToT passively waits for
victims to approach a malicious NFC tag that is embedded within
ordinary objects so that victims will not even realize that their
devices may engage in NFC communication with these tags.
That is, a ToT aims to carry out an attack without being
perceived by the victims.

% 4. さらに ToT に対して施した2つの拡張について説明
To fully explore the threat of ToT attacks, we develop a 
{\em ToT device}, which is used to mount sophisticated ToT
attacks. 
It consists of a processor, communication interface such as Wi-Fi, and
an NFC-tag emulator, which is a device that makes use of the NFC card
emulation mode and can act as multiple NFC tags.
The standard operation of a {\em ToT device} is as follows.
It first presents a malicious URL to the victim device.
The {\em ToT device} works with a web server behind the URL. The web server
fingerprints the victim device and conveys the type of the device to
the {\em ToT device}.
The {\em ToT device} uses this information to tailor additional tags
to be sensed by the victim device.

Although malicious NFC tags can induce a victim device to do certain
low-risk actions such as opening a URL without prompting the user,
higher-risk actions such as pairing with a Bluetooth device do require
user confirmation.
To deal with this problem, we develop several alternative techniques
to deceive the user into confirming the prompt.
In addition, we develop a new technique named {\em Phantom touch generator},
which aims to deceive victim devices into sensing phantom touch events
in their touch screens by applying strong electromagnetic field at a
specific frequency to trigger capacitive coupling.
These techniques enable an attacker to carry out various attacks
without being noticed by the device owner who unintentionally
puts the device close to the installed ToT.

% 5. 本研究の貢献
We make the following contributions:
\begin{mybullet}
\item We present a novel class of attacks that we call ToT, 
  which injects malicious functionalities into common objects (Section~\ref{sec:overview}). 
\item We develop the two effective techniques; ``ToT device''
  (Section~\ref{sec:sdmt}) and ``Phantom touch generator'' (Section~\ref{sec:last_touch}). 
\item We demonstrate the feasibility of ToT attacks using 24 smartphones
  for the NFC reading experiments and 7 smartphones for the 
  {\em Phantom touch generator} experiments (Section~\ref{sec:eval}).
\item We provide possible countermeasures against the threats of 
  ToT attacks (Section~\ref{sec:discuss}).
\end{mybullet}

%% % 6. 本論文の章構成
%% \noindent{\bf Roadmap.} Section~\ref{sec:background} provides background
%% information on NFC and touch screen controller, which both play a key
%% role in the attack we propose.
%% Section~\ref{sec:overview} describes the high-level overview
%% of ToT attack.
%% In Sections~\ref{sec:sdmt} and~\ref{sec:last_touch}, we introduce the two
%% techniques, software-defined malicious NFC tags and the touch
%% scatterer, respectively.
%% In Section~\ref{sec:eval} we present the feasibility study of {\em
%%   ToT} attack using the several smartphones/tablets. 
%% Section~\ref{sec:discuss} provides discussion on the feasibility of
%% the attacks with ToT, possible countermeasures against the
%% attack, and the ethical considerations.
%% Section~\ref{sec:related} summarizes the related work and
%% Section~\ref{sec:conclusion} concludes the paper with a summary. 

\begin{figure*}[tbp]
    \begin{center}
      \includegraphics[width=133mm]{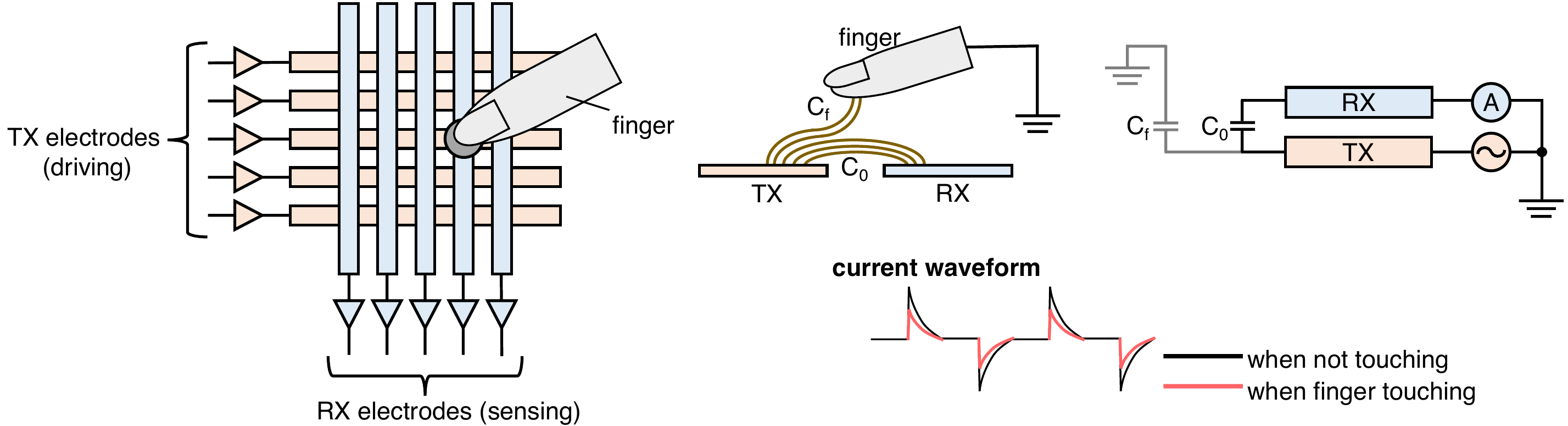}
      \caption{Touch detection mechanism of mutual capacitance touchscreen~\cite{mutual_capacitance_kyusyu}.}
      \label{fig:mutual_capacitance}
    \end{center}
\end{figure*}

\section{Background}
\label{sec:background}
In this section, we provide background information on the two key
technologies used in our attack, NFC and capacitive touchscreen, which
are widely used in smartphones.

\subsection{NFC}
% 1. 導入
Near-Field Communication (NFC) is a short-range wireless communication
technology widely used in many applications, e.g., contactless payment
systems, transit passes, smart posters, and smartphone apps.
According to Ref.~\cite{nfcphoneforecast},
the number of smartphones equipped with NFC is drastically increasing
year by year. Roughly two-thirds of all smartphones shipped in 2018
are expected to be equipped with NFC.

NFC makes use of magnetic inductive coupling to communicate between
two devices. NFC devices can be classified into two types: active and
passive NFC device. An active NFC device has its own power
source and acts as an NFC reader/writer. A passive NFC device, e.g.,
NFC-equipped IC card or NFC tag, does not have its own power source.
When an active NFC tries to read data from a passive NFC device, it
emits a weak magnetic field to induce electric current in the passive
NFC device. Given the electric current, the passive NFC device encodes
data and generate a magnetic field to induce electric current in the
active device. While the theoretical working distance of NFC is up to
20 cm, the practical working distance is a maximum of about 4 cm.

% 2. ユーザインタラクションがないこと⇒脅威の話
NFC is a communication protocol that can exchange data just by
bringing NFC compatible devices close to each other. In many NFC
applications, communication is established without going through user
interaction; e.g., mobile payments are completed just by placing the
two devices at a close distance. This design leads to high
usability. However, the high usability of NFC raises several security
issues. Although the NFC communication range is limited to only a few
centimeters and tags can be configured to be read-only, the NFC
service can be easily exploited by a simple attack — replacing the
existing NFC tag with a malicious NFC tag. Several studies have
reported the threats of malicious NFC
tags~\cite{rieback2006your,mulliner2009vulnerability,verdult2011practical,miller2012don,wallofsheep,gold2015testbed}.
We will summarize these studies in Section~\ref{sec:related}.
Wall of Sheep, an organization that makes people aware of security
risks, recommends that people should not trust NFC tags created by
third parties and take precautions~\cite{wallofsheep}.

% 3. Android で出来てしまうことを列挙
Android OS has supported NFC technology from version 2.3. Note that
Android smartphones can work as either a passive or an active NFC
device. In the following, we focus on the characteristics of Android
smartphones as an active NFC device. When an Android device is held
over an NFC tag, Android OS can perform various operations by reading
the data recorded in the NFC tag. Table~\ref{tab:nfc-android} lists
the operations that can be launched by reading an NFC tag. Recently,
Google announced a new technology called Android Instant
Apps~\cite{instant-apps}. It allows a user to use apps without
downloading/installing them. Android Instant Apps can be accessed via
a web link or an NFC tag containing the web link. Thus, reading an NFC
tag can launch a new app that has not been installed on the smartphone.

\begin{table}[tbp]
  \footnotesize
  \caption{Android OS operations that can be launched by reading an NFC tag.}
    \label{tab:nfc-android}
    \begin{tabular}{|p{43.5mm}|p{26.9mm}|}
      \hline
      \bf{operation} & \bf{requests user approval} \\
      \hline\hline
      open a specified URL & No \\
      \hline
      launch a specified app & No \\
      \hline
      send an Intent to an NFC-enabled app & No \\
      \hline
      launch an Instant app (new) & No \\
      \hline
      send email to specified address with specified subject and body
      & Yes \\
      \hline
      connect to specified Wi-Fi AP & Yes \\
      \hline
      pair with specified Bluetooth device & Yes \\
      \hline
    \end{tabular}
\end{table}

\subsection{Capacitive Touchscreen}
% 1. 導入 静電容量、相互静電容量の話。サイズなども
Majority of the current mobile devices such as smartphones and tablets
are equipped with touchscreens. While there are various technologies
for sensing touch, mutual capacitive sensors are widely used for
smartphones as they have high durability, fast response, and
multitouch support~\cite{du2016overview}.

As shown in~\figref{fig:mutual_capacitance}, a mutual capacitance
touchscreen controller consists of transmitter (TX) electrodes and
receiver (RX) electrodes,
which are mutually coupled, e.g., $C_0$ in the figure. The grid of TX and
RX electrodes is used for sensing touch events. As the human body has
a capacitance, it can act as a capacitor. When a finger approaches to
the screen surface, it passes electric charge onto the touchscreen
sensors through mutual capacitance ($C_f$ in the figure). Thus, the
touchscreen controller can detect touches by measuring the changes in
electric current that flows into the RX electrodes; the current
changes are caused by the changes in capacitance between the TX and RX
electrodes. The pair of TX and RX electrodes for which the changes are
detected is used to locate the area of touch.

% 3. タッチスクリーンに対する脅威
It is known that a touchscreen controller in a smartphone can
malfunction due to noise signals leaked from the smartphone's battery charger
or screen~\cite{cypress_noise_immunity}. Touchscreen controller
manufacturing companies have developed countermeasures against the
electromagnetic interference (EMI) caused by noise signals, which are
relatively weak. However, when a stronger noise signal is
intentionally applied to a touchscreen controller, false touch events
can be generated. As some hobbyists have reported~\cite{youtube_plasma_goes_crazy,youtube_plasma_touch_from_back}, it is known
that false touch events occur when a smartphone is brought close to a
commercial plasma ball, which is powered by an oscillator and a
high-voltage transformer circuit producing a large alternating
voltage, typically 2–5 kV at around 30 kHz~\cite{campanell2010,plasma_ball}. 
The strong electric field generated by the electric circuit of the
plasma ball causes capacitive coupling with the touchscreen sensors;
the coupling causes changes in electric current flowing into the RX
electrodes and the changes are detected as random touch events.

%% Inspired by this observation, we propose a new attack, named 
%% {\em Phantom touch generator} in Section~\ref{sec:last_touch}. The key
%% idea is to build a system that intentionally generates a strong
%% electric field with a specified frequency to cause false touch events
%% on a victim's smartphone.

\section{Trojan of Things}
\label{sec:overview}
In this section, we present the overview of ToT attacks. We first
describe our threat model. We then introduce several attacks using
malicious NFC tags. Finally, we present examples of ToT
implementations and their implications.

\subsection{Threat model}
In this work, we assume an attacker has embedded a malicious NFC
device into a targeted thing in advance.
If the target is a small and portable thing such as banknote or
clothing, the attacker embeds a malicious NFC tag into it.
This device can carry out a simple attack.
If the target is a large and stationary thing such as a table,
the attacker can embed several components, e.g., an NFC-tag emulator,
a single-board computer, and high-voltage transformer, in it.
This is what we call a {\em ToT device}, which is used to carry out
sophisticated attacks.

We also assume the victim has an Android smartphone equipped with
NFC. The victim unintentionally places the smartphone close to a ToT,
and the smartphone automatically reads a malicious NFC tag/emulator
when it is unlocked and not in the sleep mode. The validity of this
assumption will be discussed in Section~\ref{sec:discuss}.
After reading the NFC Data Exchange Format (NDEF) records stored in
the malicious NFC tag/emulator, the smartphone will execute a
corresponding operation used for attacks, which will be described in
the next subsection.

\begin{figure}[tbp]
  \centering
  \includegraphics[width=80mm]{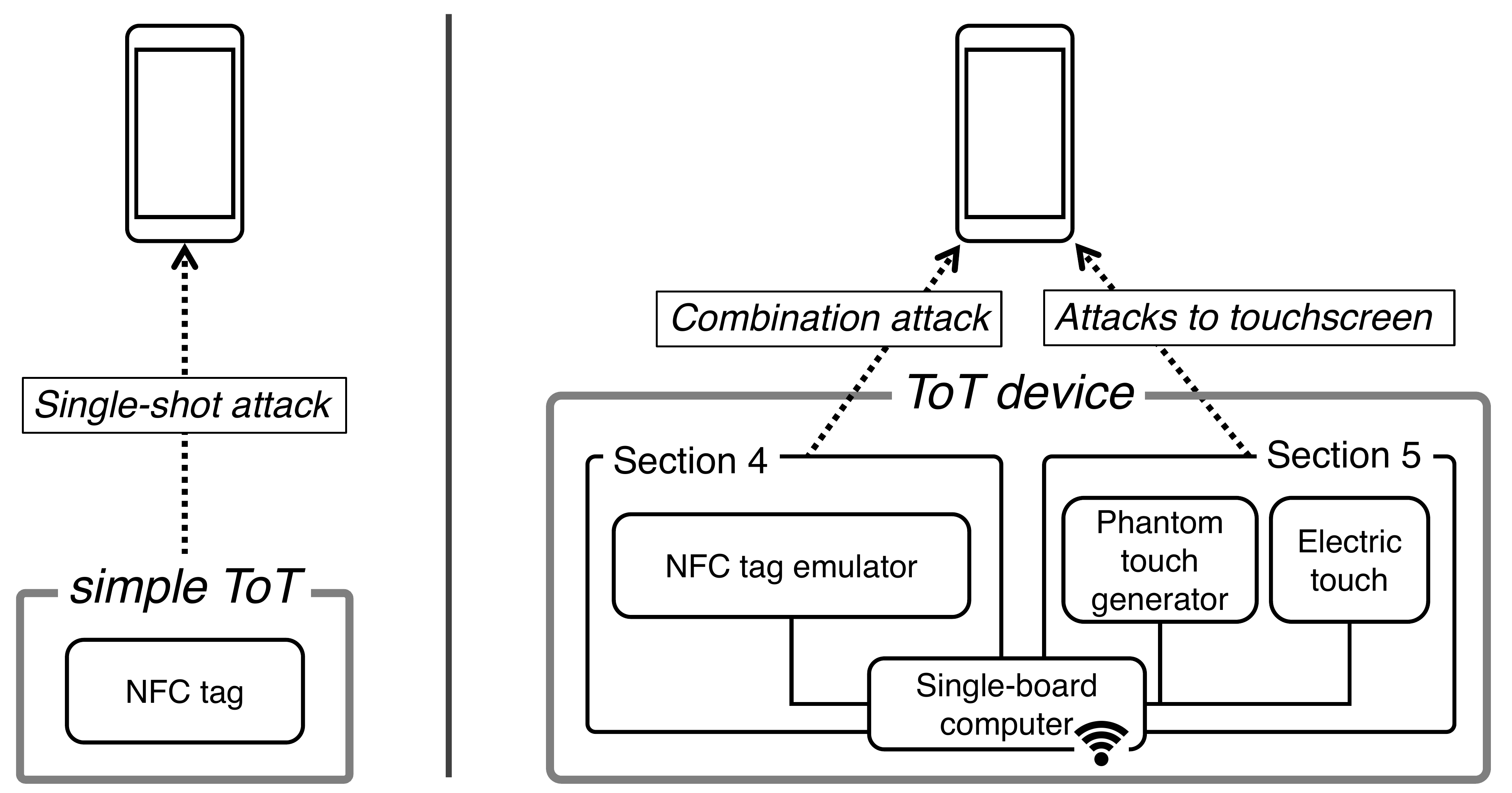}
  \caption{Overview of ToT attacks.}
  \label{fig:tot-overview}
\end{figure}

As triggering high risk actions such as connecting to Wi-Fi AP
requires user approval by displaying a dialog box with a confirmation
message, we develop two techniques to evade the user approval process.
The first is to 
mislead the user into approving the dialog box by different ways of
manipulating the UI such as showing a deceptive message or dimming
relevant parts of the display (Section~\ref{sec:sdmt}).
The second technique is an attack on the touchscreen named 
{\em Phantom touch generator} (Section~\ref{sec:last_touch}).
Figure~\ref{fig:tot-overview} summarize the types of ToT, possible
attacks, and the system components.

%% As shown in Fig.~\ref{fig:tot-overview}, this attack requires
%% embedding of several equipments into the ToT; thus, the ToT needs to
%% be large and stationary, such as a table.

%% \begin{table}[tbp]
%%   \centering
%%   \caption{Summary of ToT attacks. A1: single-shot attack,
%%     A2: combination attack with software-defined malicious NFC tags,
%%     and  A3: touch scatter.}
%%   \label{tab:tot-summary}
%%   \begin{tabular}{l|l|l}\hline
%%     Mode  &  Examples & Attacks \\\hline\hline
%%     single tag & banknotes, clothings & A1 \\
%%     % middle & an attache case, a flower vase & a NFC tag emulator,  a single-board computer & A1, A2 \\
%%     ToT device & tables, desks & A2, A3  \\
%%     \hline
%%   \end{tabular}
%% \end{table}

\subsection{Attacks using the malicious NFC tags}
As shown in Table~\ref{tab:nfc-android}, two types of operations can
be invoked via NFC:
operations that require user approval and operations that do not. 
The latter require will be automatically
executed if an NFC tag is brought close to a smartphone. We call an
attack that makes use of such operations as a {\em single-shot
  attack}. A representative example of a single-shot attack is opening
a malicious URL in a browser; such a malicious website can trigger
download/installation of a malware on the smartphone~\cite{wallofsheep}.

By combining multiple single-shot attacks, we can create more
sophisticated attacks, which we call {\em combination
attacks}. {\em ToT device} is a system that implements combination attacks.
Combination attacks enable an attacker to establish device
fingerprinting. As shown in Section~\ref{sec:sdmt}, device fingerprinting is useful to
infer the language used for the device; the information can be used
to display a dialog box with a deceptive message to the victim.
The fingerprint information can also be used for displaying a
dialog box with a suitable message, which needs to be adaptive to
the vendor-specific customization of confirmation message strings.

Operations that require user approval can be used for high risk
attacks. For instance, by forcing a device to connect to a malicious
Wi-Fi AP, the attacker can establish the man-in-the-middle attack. Or,
the attacker can even take complete control of the smartphone by
forcing the device to pair with a Bluetooth mouse, which can be used
as a remote control. Thus, evading the user approval process is a key
success factor of the attacks. One way to evade the use approval
process is to display a dialog box with a deceptive message, as we
discussed above. Actual examples of composing such a deceptive message
will be described in Section~\ref{sec:sdmt}. Another way is to employ
the new attack we developed, Phantom touch generator, which will be
described in Section~\ref{sec:last_touch}.

\subsection{Examples of ToT implementation}
\label{sec:hidden_nfc_things}
In order to let a victim accidentally scan a malicious NFC
tag/emulator on his/her smartphone, a malicious NFC tag/emulator
should be embedded in a thing that has many opportunities to come
close to a smartphone. In this section, we present two examples of a
simple ToT and an example of a {\em ToT device}.

\begin{figure}[tbp]
  \begin{center}
    \begin{minipage}[b]{0.48\columnwidth}
      \begin{center}
        \includegraphics[width=30mm]{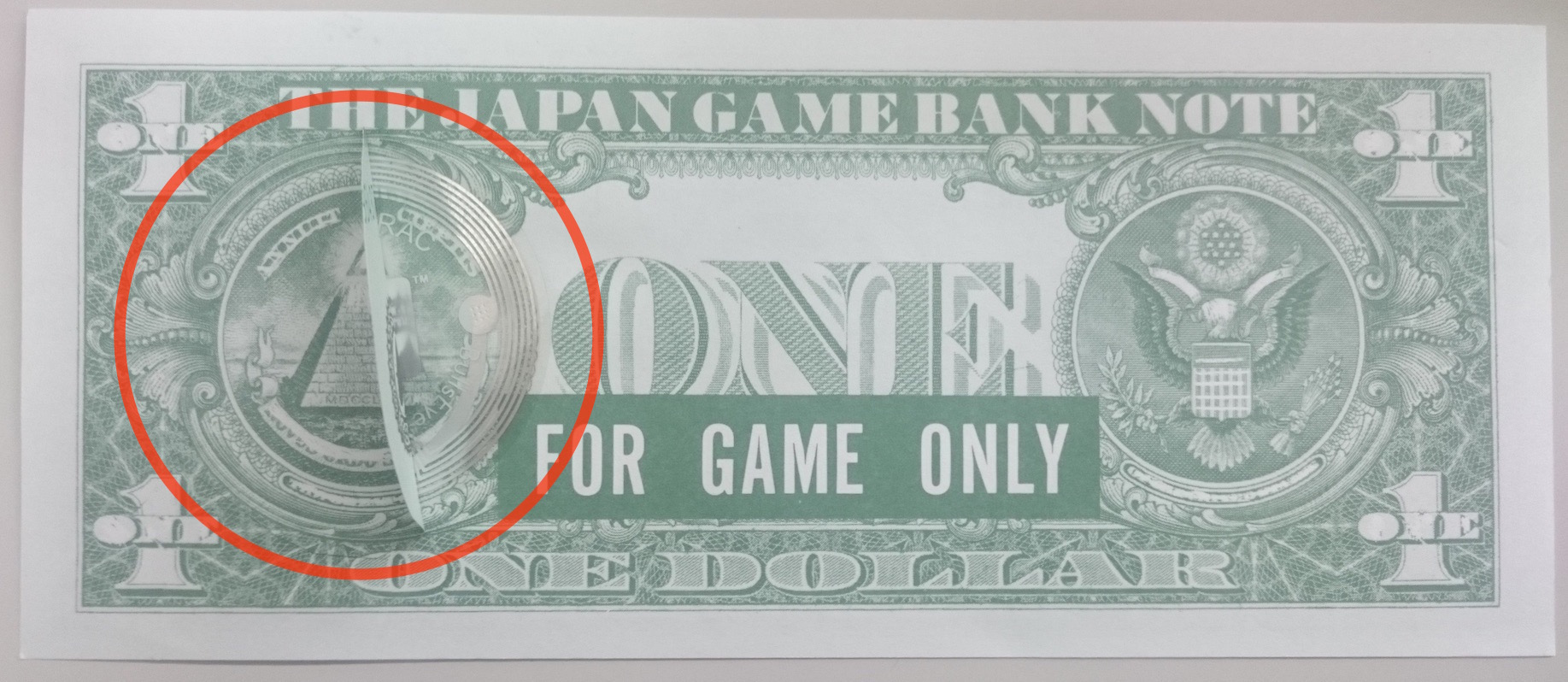}
      \end{center}
      \vspace{-5mm}
      \caption{ToT as a\newline banknote.}
      \label{fig:money}
      \vspace{2mm}
    \end{minipage}
    \begin{minipage}[b]{0.48\columnwidth}
      \begin{center}
        \includegraphics[width=30mm]{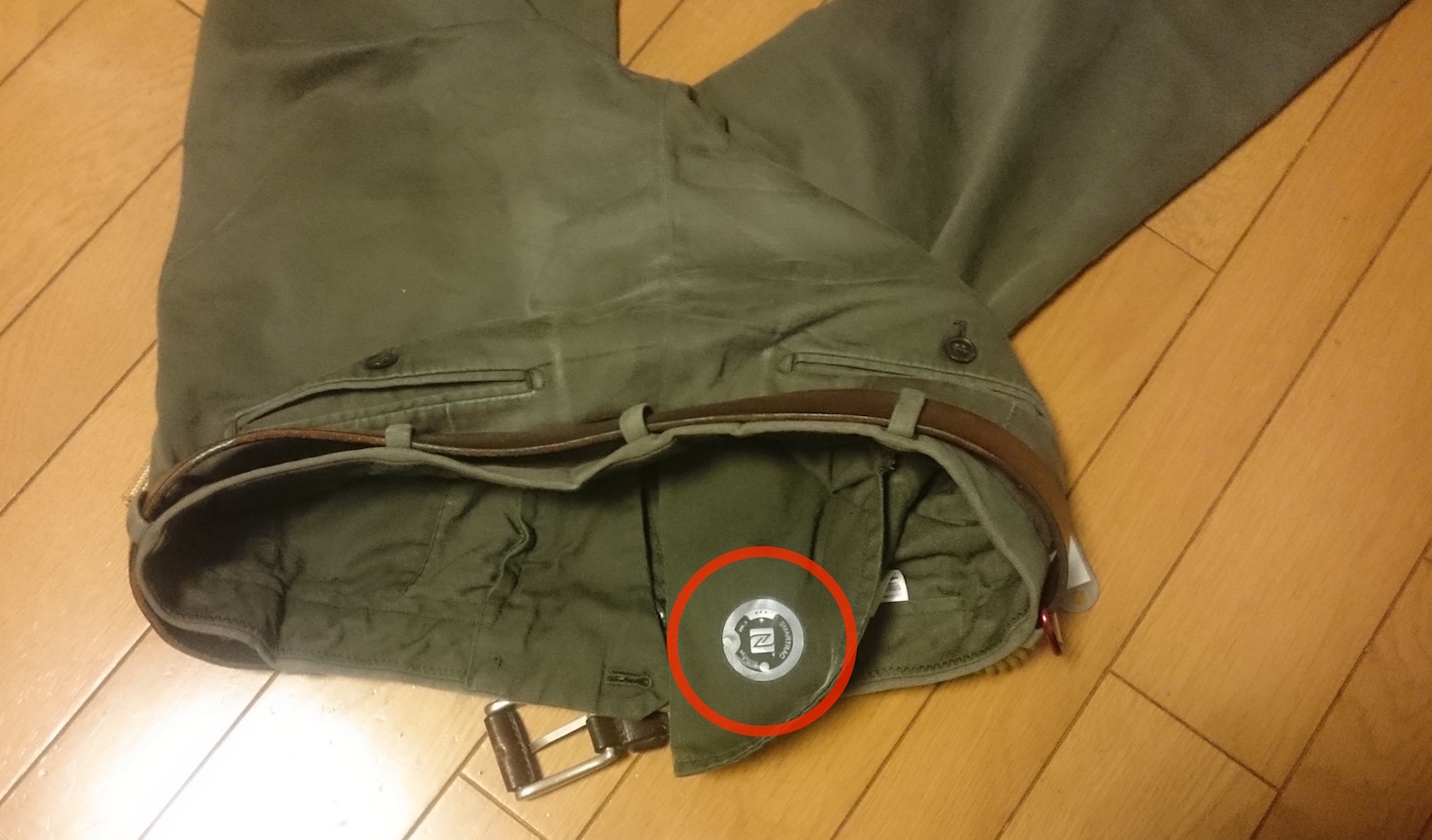}
      \end{center}
      \vspace{-5mm}
      \caption{ToT as a pair of trousers.}
      \label{fig:trousers}
      \vspace{2mm}
    \end{minipage}
    \includegraphics[width=80mm]{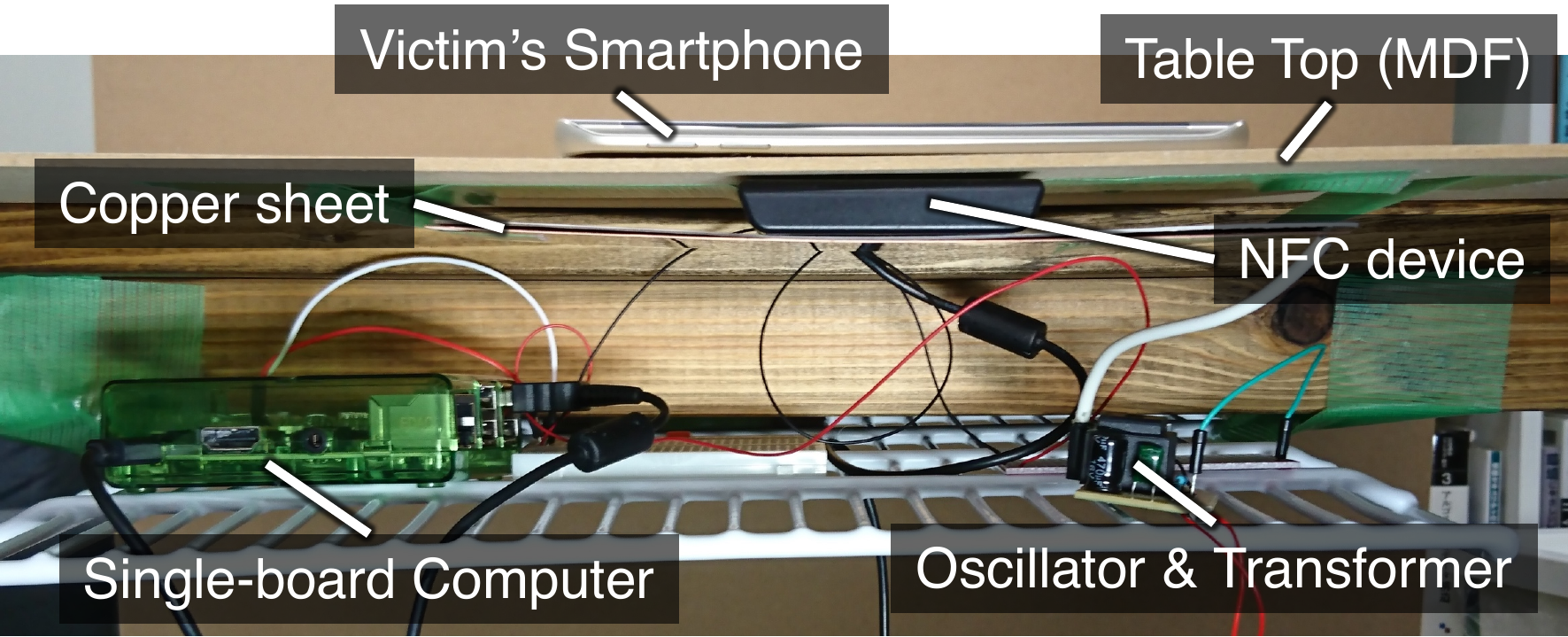}
    \vspace{-5mm}
    \caption{{\em ToT device} as a desk.}
    \label{fig:desk}
  \end{center}
  \vspace{-5mm}
\end{figure}

Figure~\ref{fig:money} presents an implementation of a simple ToT in a
banknote. We embedded a malicious NFC tag into a toy banknote imitating
a one dollar bill. The NFC tag is embedded into an area indicated by
a circle, as shown in the figure. We placed the bill in a wallet and
the wallet in the pocket. When an unlocked smartphone was placed in
the pocket, the smartphone read the data from the malicious NFC
tag. Thus, the ToT attack is easy to deploy and feasible. The
implications of such a ToT attack are as follows. The most
notable feature of banknote is that it physically circulates from
person to person. Therefore, by embedding a malicious NFC tag in a
banknote, several smartphones can be attacked one after another. In
addition, since many people may carry their wallet and smartphone
together in their pockets or bags, there are many opportunities for
the ToT to attack an individual smartphone. We also note that it is
not easy to track the attacker once a mobile ToT is disseminated into
the real world.

Figure~\ref{fig:trousers} presents an implementation of an another
simple ToT in a pair of trousers. A malicious NFC tag is embedded into
an area surrounded by a circle shown in the figure, i.e., on the back
of the pocket. Since clothes may be washed in a washing machine,
sewing a durable (e.g., laminated) NFC tag is suitable for this
attack. When an unlocked smartphone was placed into a pocket with a
malicious NFC tag, the smartphone successfully read the data from the
malicious NFC tag. The implications of such a ToT attack are as
follows. The target can be extended to various clothing items such as
clothes displayed at a clothing retailer, rental clothes, laundary
being dried outdoors, or a suit hanging on a chair. Since clothing is 
personal, it can also be used for a targeted attack. By
embedding a malicious NFC tag in a part which has a high possibility
of being close to a smartphone, such as a chest pocket, a trouser
pocket, or the end of a sleeve, we can increase the opportunities for
a smartphone to read the malicious NFC tag.

Finally, Figure~\ref{fig:desk} presents an implementation of {\em ToT
device} using a desk. In this implementation, an NFC emulator, a
single-board computer, and other devices needed for Phantom touch
generat are embedded under the table top of the desk. We will evaluate
how the thickness of the table top of the desk affects the NFC reading
in Section~\ref{sec:eval}. The {\em ToT device} will be
installed at a fixed location such as in a library.
If the attacker places the {\em ToT device} at a public space, a large
number of person will come close to it.

\section{ToT Device}
\label{sec:sdmt}

In this section, we first provide an overview of {\em ToT device}.
We then present combination attacks, which can be established using
the {\em ToT device}.

\begin{figure}[tbp]
  \centering
  \includegraphics[width=80mm]{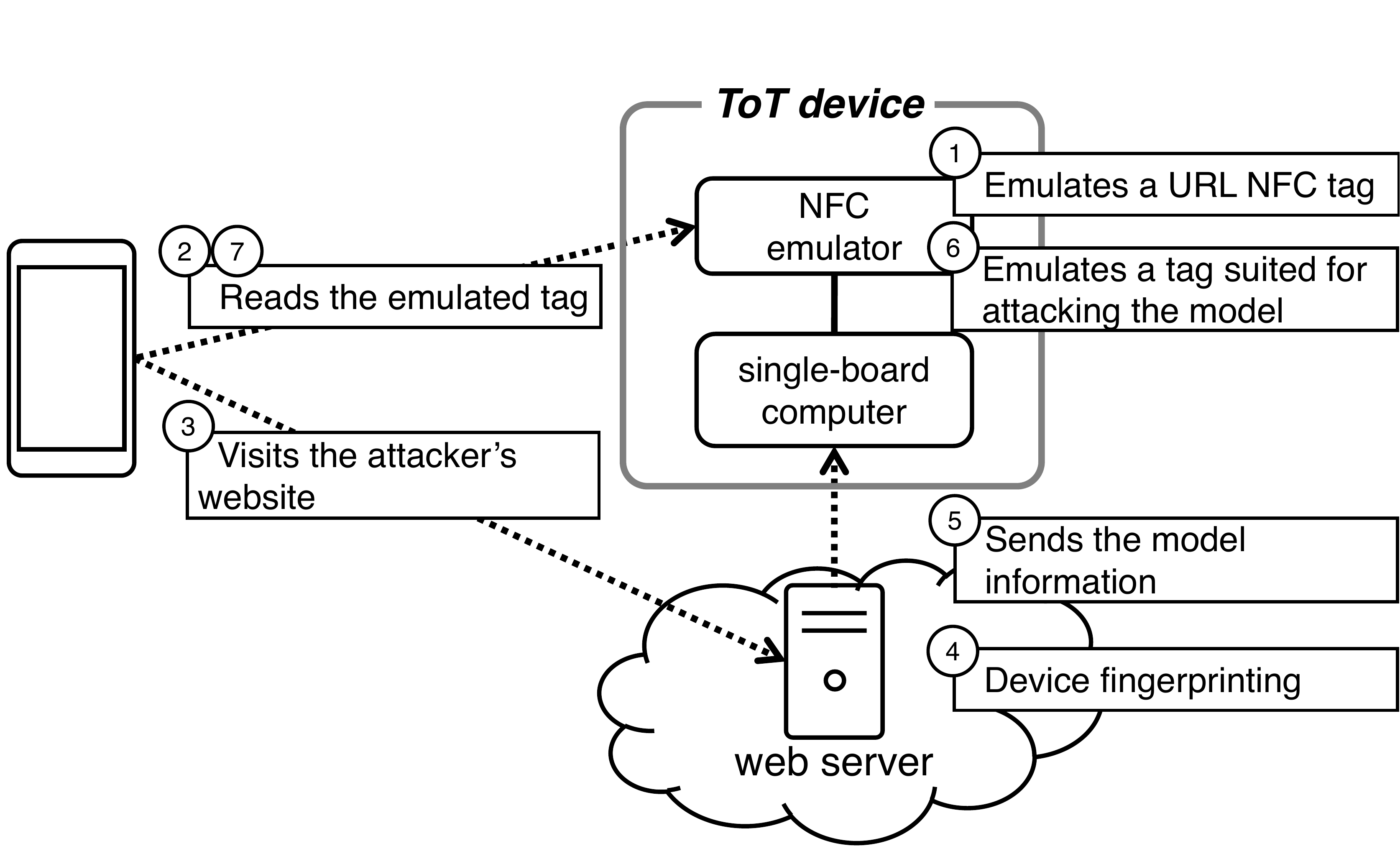}
  \caption{Overview of {\em ToT device}.}
  \label{fig:sdmt-overview}
\end{figure}

\subsection{Overview}
Figure~\ref{fig:sdmt-overview} presents an overview of the
{\em ToT device}.
It comprises the two primary components, an NFC tag emulator and a
single-board computer with a Wi-Fi controller installed.
The {\em ToT device} works with a web server, which can be set
anywhere connected to the Internet, e.g., a cloud server. 
%% If the attacker can assign a global IP address to the single-board
%% computer, the web server can be installed on the single-board
%% computer. 
We note that the attacker also needs to install a power source.
As an NFC-tag emulator, we used Sony RC-S380 for our experiments.
By using the NFC-tag emulator, we can dynamically switch the NFC tags
according to the attack scenario.

We now describe how the {\em ToT device}
works using the example shown in the figure.
\begin{mybullet2}
\item[\ctext{1}] First, the NFC tag emulator acts as an NFC tag with a URL data
recorded and waits for a victim to approach.
\item[\ctext{2}] When the victim's smartphone comes close to the ToT, it reads the tag
and launches a browser to open the URL.
\item[\ctext{3}] The browser then connects to the website specified by the recorded URL.
\item[\ctext{4}]
The website employs device fingerprinting by using
JavaScript to collect information about the victim's device.
\item[\ctext{5}] The website sends the device fingerprinting information to the
computer onboard the ToT device.
We assume that the computer has Internet access.
\item[\ctext{6}] Upon receiving the device information, the
  computer determines the tag suited for the victim's
  device and rewrites the NDEF record of the NFC tag emulator.
\item[\ctext{7}]
Finally, the victim's smartphone reads the new NDEF record from the
tag and gets attacked again.
Note that the smartphone will read a new record after the
emulator is turned off (which implies that the old tag went away) and
turned on again. 
\end{mybullet2}

\subsection{Combination attacks}
\label{sec:combination}

As we have shown, the framework of software-defined malicious NFC tags
enables an attacker to employ the device fingerprinting.
By using the device fingerprinting information, the attacker can
further perform a {\em targeted attack}, which leverages
the intrinsic features of the mobile devices, e.g., language setting, vendor
customization, and the noise tolerance characteristics of the touchscreen
controllers, etc.
In the following, we present the two applications of the combination
attacks -- {\em deceptive message trap} and
{\em exploiting installed apps}.
Both attacks aim to deceive a victim into touching a button, which
establishes the attack, e.g., connecting to a malicious Wi-Fi AP that
employs the man-in-the-middle attack.

\subsubsection{Deceptive message trap}
\label{sec:deceptive}

%% WiFiConfigレコードやBTSSPレコードによる攻撃では，OSが設定変更前に確認ダイアログを表示し，ユーザの承認を求める．
%% この確認ダイアログによるOSとユーザの対話を妨害し，
%% ユーザの判断をミスリードすることで攻撃が成功しやすくなると考えられる．
To make the descriptions easy to follow, we first describe a case
where the attacker does not use the device fingerprinting. We then
describe a case where the attacker needs device fingerprinting.

The deceptive message trap is an attack that aims to deceive a
victim into touching a button that establishes the attack.
We focus on the scenario of a Wi-Fi attack as a
representative example.
In this scenario, the goal of the attacker is to deceive a victim
in touching the ``CONNECT'' button when a modified message pops up
after reading the malicious NFC tag with the WiFiConfig record.

In the Android OS, as of February 2017, the format of the confirmation
message invoked by the WiFiConfig NFC record is defined in the
file named,
\url{android/platform/packages/apps/Nfc/res/values/strings.xml}~\cite{ssid_string}.
\figref{fig:ssid_string} summarizes an excerpt of the main part.
Here, the strings shaded with gray are replaced with the
service set identifier (SSID) value specified in the WiFiConfig NFC
record.
SSID is an identifier for a Wi-Fi AP.
Since the maximum length of the strings used for specifying a SSID is
set to 32 bytes and the SSID encoding scheme allows the use of the
UTF-8 charset~\cite{IEEE_Std}, the attacker can tweak the SSID strings
to deceive a victim.

We show an attack scenario using this trick.
The attacker creates a malicious NFC tag with the SSID of WiFiConfig
record set to ``again''.
When the victim's smartphone approaches to the {\em ToT device} with the
malicious NFC, the following confirmation message pops on the
screen:\\
\quad\quad\quad\quad\quad\ovalbox{\em{Connect to network again?}}\\
When the victim notices this message popping up, she/he may think that the
Internet connection is lost and the smartphone is asking
to reconnect to the previously connected network, and will touch
``CONNECT''. Thus, the man-in-the-middle attack is established. Note
that a single-board computer can work as a malicious Wi-FI AP. Along
this line, the attacker can create various deceptive messages such as\\
\quad\quad \ovalbox{\em{Connect to network to prevent the data lost?}}\\
Such a message will threaten the victim into touching the
``CONNECT'' button, which again will connect the smartphone to the
malicious Wi-Fi AP.

We now turn our attention to the case where the attacker needs device
fingerprinting. As the format shown in~\figref{fig:ssid_string} represents the case
for uncustomized Android with the language configured to English, the
attacker may want to customize the message according to the language
used by the victim and the model of the smartphone.
Through the analysis of 24 Android smartphones equipped with NFC, we
found that several vendor customizations use different formats for the
confirmation messages.
For reference, we summarize the result in
Table~\ref{tab:result_studies},~\ref{tab:messages_wifi},
and~\ref{tab:messages_bt} (Appendix).
To cope with such differences, the attacker can use the information obtained from device
fingerprinting, which was presented in Section~\ref{sec:combination}.

\begin{figure*}[tbp]
%  \scriptsize
  \begin{framed}
    \begin{lstlisting}
      <!-- Title for dialog where user confirms that they want to connect to the network on the tag they tapped-->
      <string name="title_connect_to_network">Connect to network</string>
      <!-- Message prompt asking the user if they wish to connect to the given network. Contains the network name (ssid).-->
      <string name="prompt_connect_to_network">Connect to network |\colorbox{gray!30}{\textless xliff:g id="network\_ssid"\textgreater \%1\$s\textless /xliff:g\textgreater}|?</string>
    \end{lstlisting}
  \end{framed}
  % <xliff:g id="network_ssid">
  \caption{Format of the confirmation message invoked by the WiFiConfig NFC record (English).}
  \label{fig:ssid_string}
\end{figure*}

%% \begin{figure}[tb]
%%     \begin{center}
%%         \includegraphics[width=100mm]{figs/ssid_spoof.png}
%%     \end{center}
%%     \caption{Prompt before connecting to the network with SSID ``再接続$\backslash$nインターネット'' (Nexus 7).}
%%     \label{fig:ssid_spoof}
%% \end{figure}

%% ただし，スマートフォンメーカーによるAndroid OSの改変により，表示される
%% メッセージの定義が\figref{fig:ssid_string}と異なる端末が存在する．
%% 例としてXperia~Z3（\tabref{tab:equipment}）で表示される確認ダイアログを\figref{fig:ssid_xperia}に示す．
%% メーカーによるメッセージ改変の実態については\ref{sec:experiment}章で詳細に述べる．
%% メッセージの定義の差異に対応するには，まず\ref{sec:mal_webpage}項で述べた方法で機種を特定し，
%% 次に\ref{sec:impl_sequential}節で述べた手法を用いて，機種に合わせた
%% WifiConfigレコードを読み込ませればよい．

\subsubsection{Exploiting Installed Apps}
This attack leverages the apps installed in the victim's
smartphone. For this attack, the attacker specifies
``Android Application`` in the NDEF record of a malicious tag. After
reading the application tag, Android OS will automatically execute the
application specified in the record without requiring user approval.
There are two variations of this attack. The first variation aims to
make the deceiving message look real by intentionally creating a
context. The attacker first sets an Application NFC tag that launches
a popular SNS app such as Facebook. Subsequently, the Facebook app
appears on the screen of the victim's smartphone. The attacker then
sets the WiFiConfig NFC tag using the technique described in the
previous subsection. The message popping on the screen appears as
follows:\\
\quad\ovalbox{\em{Connect to network ? Facebook app is requesting.}}\\
Since the dialog box of this message appears on top of the Facebook
app, it looks as if the message is originated from the Facebook
app. Some Facebook users may touch the
``CONNECT'' button, never knowing that the message is for connecting
to a malicious Wi-Fi AP. Note that to create this message, we set the
following text string as the SSID:  ``$\backslash$u202E.gnitseuqer si ppa
koobecaF``, where `$\backslash$u202E' is a Unicode character known
as RIGHT-TO-LEFT OVERRIDE.

Another variation is to make use of a utility application that adjusts
the brightness of the screen, e.g.,
``Screen Filter''~\cite{screenfilter}, which has
been installed by more than 5 million users as of February 2017. Since
the aim of such applications is to reduce eye stain while using the
smartphone during nighttime, the users usually adjust the brightness
level lower than the default setting. Therefore, when the app is
executed, the screen gets darker, which makes the characters displayed
on the screen difficult to read during daytime. The attacker first
sets an application tag that executes such an app. If the victim's
smartphone comes close to the tag, the screen automatically becomes
darker. The attacker then switches the tag to the Wi-Fi tag mode. A
pop-up message that is difficult to read during daytime automatically
appears on the dark screen. The users may accidentally click the
``CONNECT'' button in such a situation.

We provide screenshots of the attacks described above in the
appendix.

\section{Attacks to Touchscreen}
\label{sec:last_touch}

In this section, we will first describe the new attack, named
{\em Phantom touch generator}, which aims to alter the selection of a
button on a screen; i.e., while a victim thought that she/he touched
the button ``A'', the attack can scatter the recognized touched
position and make the operating system recognize another button ``B''
touched. We then present another attack to the touchscreen; an
attacker installs a circuit board on top of the table/desk, which can
directly cause touch events at an arbitrary position.

\subsection{Phantom touch generator}
\label{sec:touch_flood_attack}

\subsubsection{Overview}
Phantom touch generator is an attack that aims to scatter touch events
around the original touch area; i.e., even though a victim touches a
``CANCEL'' button, which should cancel the request to connect to a
malicious Wi-Fi AP, the attack make the operating system recognize the
event as a touch of another button, ``CONNECT,'' in a probablistic
way.
Thus, the attack can trick the user, with a certain success rate.
In the following section, we aim to present the basic mechanism of
Phantom touch generator and reveal the conditions that are needed to
establish the attack.

The key idea of Phantom touch generator is to intentionally cause the
malfunction by injecting intentional noise signals from the
external. As we discussed in Section~\ref{sec:background}, a
touchscreen controller mounted on a smartphone can cause a malfunction
due to the noise signals~\cite{cypress_noise_immunity}.
The malfunctions include three types: (1) ``false touch,'' which
reports touches at positions where no touch is present, (2)
``no-touch,'' which reports that a touch does not exist
when a finger touches the area, and (3) ``jitter,'' which reports
the coordinates distributed around the true touch point~\cite{cypress_noise_immunity}.

As we had hints from the experiments of toy plasma
balls~\cite{plasma_ball}, we found that we can intentionally cause the
malfunction by generating an electric field near the capacitive
touchscreen controller, using an electric circuit that can produce
large alternating voltage. Applying such signals using a metal plate
can create a capacitive coupling with the capacitive sensors of
touchscreen controllers. As we showed in
Section~\ref{sec:background}, the capacitive coupling causes the
changes of capacitance between the TX and RX electrodes of the
touchscreen controller, and the changes will be detected as the
(false) touch events.

\subsubsection{Experiments}
To study the conditions that can cause the ``false touches,'' we
conduct several experiments using the touchscreen controller that
provides raw data collected from the capacitive sensors. In the
following, we first describe our experimental setup. Second,
we attempt to specify the intrinsic frequency of injected noise signal
to maximize the false touches. We then analyze the spatial patterns of
the false touch events on the screen with a noise injected at a
specific frequency. Finally, we study how an actual touch event by a
user affects the spatial patterns of the false touch events. This
final experiment will reveal the mechanism of Phantom touch generator.

\begin{figure}[t]
  \begin{center}
    \includegraphics[width=80mm]{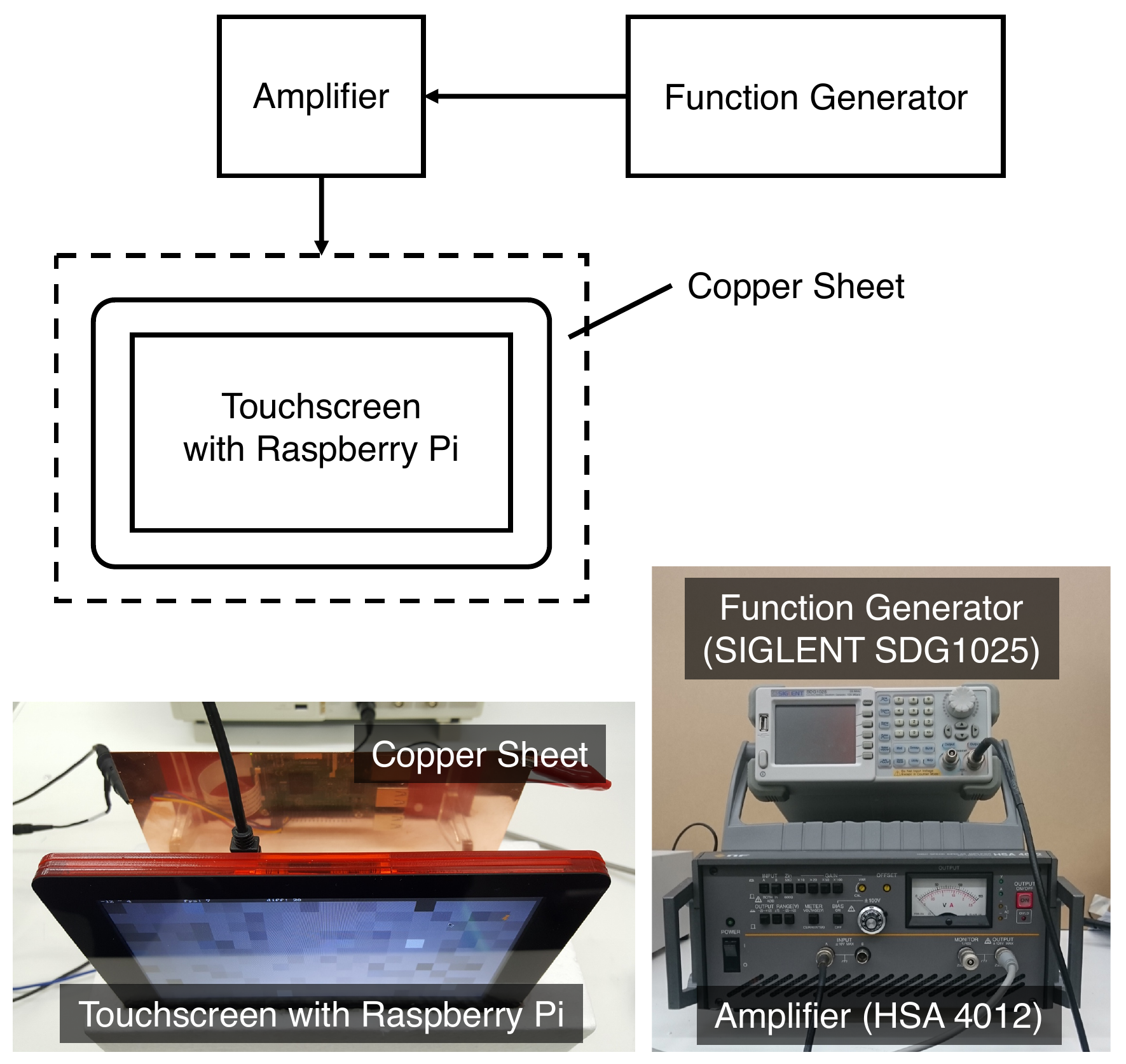}
    \caption{Experimental setup.}
    \label{fig:whitebox_setup}
  \end{center}
\end{figure}

\paragraph{Experimental setup}
\figref{fig:whitebox_setup} shows our experimental setup. Our objective
is to measure the effect of noise signals on the behavior of
touchscreens. For this experiment, we use the Raspberry Pi
7-inch Touchscreen Display. As an intentional noise signal, we use the
sine-wave signal generated by a function generator. We set a copper
sheet parallel to the touchscreen controller. This copper sheet is
used to create a capacitive coupling with the capacitive sensors. The
distance between the sheet and controller was set to 7 cm. We note
that the attack can be applied from the rear side of a touchscreen
controller, i.e., the rear side of a smartphone.

\begin{figure}[t]
  \begin{center}
    \includegraphics[width=80mm]{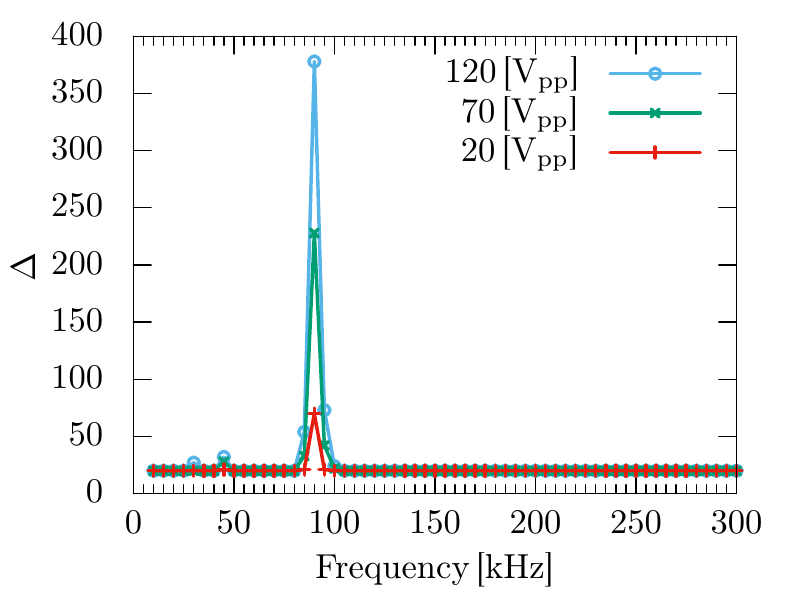}
    \caption{The effect of different frequencies on touchscreen.}
    \label{fig:rasp_frequency}
  \end{center}
\end{figure}

\begin{figure*}[tbp]
  \centering
  \includegraphics[height=39mm]{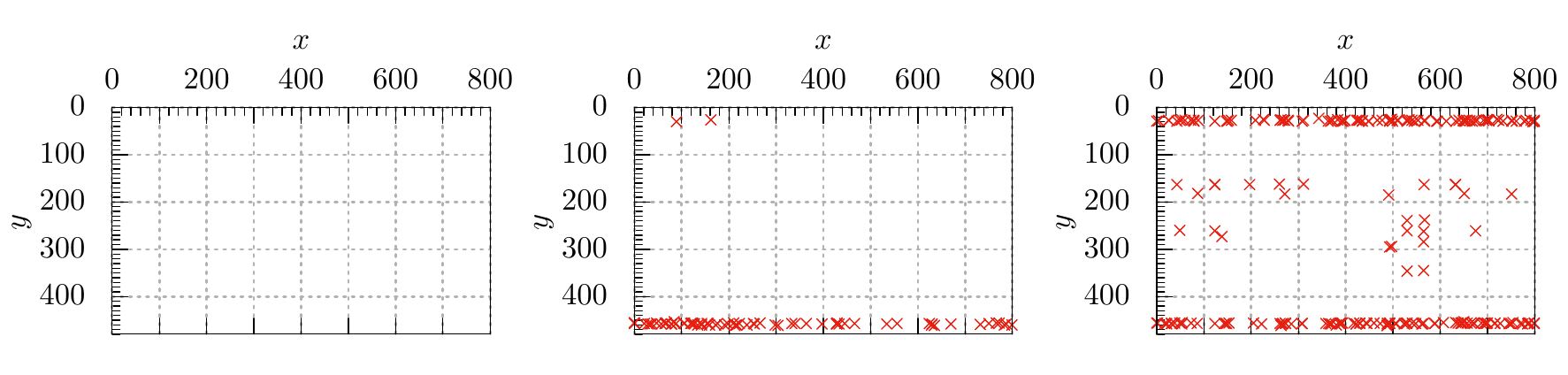}
  \caption{Coordinates of the touch points reported by
    the touchscreen controller. The injected signals had three
    different voltage values. The frequency was set to 90 kHz.
    \textbf{Left:} 20 Vpp, \textbf{Center:} 70 Vpp, \textbf{Right:}
    120 Vpp}
    \label{fig:rasp_coordinates}
    \includegraphics[height=39mm]{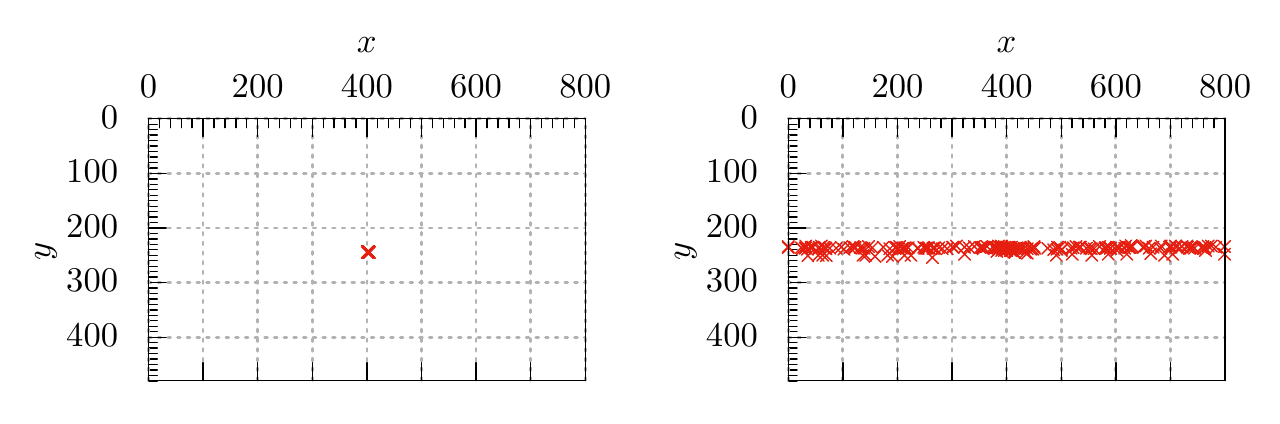}
    \caption{Coordinates of the touch points reported by the
      touchscreen controller. While the experiment a finger keeps
      touching the point centered on the screen.
      \textbf{Left:} no signal is applied.
      \textbf{Right:} a signal with 20 Vpp and 90 kHz is applied.}
    \label{fig:rasp_touch_notouch}
\end{figure*}

\paragraph{Effect of the frequencies and voltage values}
We generate sine-wave noise signals with different frequencies and
voltage values. 
We record raw capacitance values and touch events using the software
we developed. 
%, made by modifying an OSS~\cite{rasp_ft5406_capative}.
Since the touchscreen has 264 capacitance sensors, which consists
of a $12\times 22$ matrix, we can obtain 264-dimensional time-series data. 
This setup enables us to analyze the spatial patterns of the generated
touch events. 

To measure the interference intensity on the touchscreen,
we introduce a metric, $\Delta$, defined as follows.
\begin{eqnarray*}
  \label{eq:range}
  \delta_i &=& x_i - \bar{x_i}  \\
  \Delta &=& \max_i(\delta_i) - \min_i(\delta_i),
\end{eqnarray*}
where $x_i~(i\in\{1,\ldots,264\})$ is a measured value for each sensor
and $\bar{x_i}~(i\in\{1,\ldots,264\})$ is a
measured value for each sensor when noise is not injected, respectively. 
We note that $x_i$ is variable of time; our capacitance logger sampled
the raw values at the rate of 7 times per second.
In contrast, $\bar{x_i}$ was set as a static value, which was collected
when no signal was injected.
If no noise signal is applied, $\Delta$ becomes roughly 20 when there
are no touch events on the screen and $\Delta$ becomes greater than
250 when a finger touches the screen. Thus, the metric $\Delta$ can
measure the impact of noise interference. 

We measured $\Delta$, applying noise signal to the copper sheet
with three different voltages (20 Vpp, 70Vpp, and 120Vpp) and
frequencies, ranging from 5 kHz to 300 kHz. 
\figref{fig:rasp_frequency} shows the results. 
We first notice that there are clear peaks at the frequency of 90
kHz. 
This result indicates that there is a characteristic frequency of
noise that can affect the touch controller.  
As we will study in the next section, this frequency differs for 
different models of touchscreen controllers. 
So, specifying the model of the target is crucial to succeeding in the
attack. As we have seen, the device fingerprinting technique can be
used for this purpose.
We also notice that the effect of noise becomes larger with higher
voltage in the signals.
As we shall show in the next section, we need to apply higher voltage
to succeed in attacks to the smartphones.

\paragraph{Spatial distribution of the false touch events}
We now study the positions of the touch events caused by the noise 
signals.
In this experiment, nothing touches the screen. 
Using our monitoring software, we record  touch positions for 30
seconds with the sampling rate of two samples per second.
The touchscreen has an 800$\times$480 resolution and supports a
10-point multi-touch. The touchscreen controller is capable of
reporting up to 10 positions per sample.
Note that the touch events are collected from the outputs of
the touchscreen controller, not from an operating system. 

We used three different voltages (20 Vpp, 70 Vpp, and 120 Vpp) and the
following two representative frequencies: 60 kHz as a frequency not
affecting $\Delta$ and 90 kHz as a frequency affecting $\Delta$ the most.
As expected, the touchscreen does not report any touch events with the
60 kHz frequency. In the followings, we omit the results of 60 kHz
frequency. 
\figref{fig:rasp_coordinates} shows the results for 90 kHz frequency.
First, we notice that the touchscreen controller did not recognize
touch events when the voltage was set to 20 Vpp. 
We also see that higher voltage signals cause false touch events more
frequently. Second, we see intrinsic spatial patterns of touch events,
i.e., they linearly spread out on the screen\footnote{As we will
see in the next section, the direction of the spread patterns
differs for different models of touchscreen controllers; i.e.,
horizontal spread or vertical spread.}. We also see that many 
touch events are focused on
the top or bottom edges of the screen panel. 
These observations indicate that even if an attacker waits for a long
time, it seems unlikely that a false touch is fired at target
coordinates with a high probability, given the skewed spatial
distribution.

%このため、プラズマボールの信号(~\figref{fft_plasma})のように、電圧の高い、複数の周波数成分を含む信号は、多くのTouchscreenに対して、
%影響を及ぼすと思われる.
%\begin{figure}[t]
%    \begin{center}
%        \includegraphics[width=70mm]{figs/black_fft.BMP}
%        \caption{FFT Result of the Plasma Ball's Signal}
%        \label{fig:plasma_fft}
%    \end{center}
%\end{figure}

\paragraph{Limiting the dispersion with a real touch event}
After several trials, we found that touching on a screen can fix the
skewed spatial distribution of false touches. Although not conclusive
due to the ``black box'' nature of the touchscreen controllers, we
conjecture that the touching with a finger stabilizes the area of
capacitive coupling. The good feature of this phenomenon is that while
touching on a screen makes the distribution focused on a certain area,
it still keeps scattering the touch events; thus, it can create false
touch events in a more predictable way.

We repeated the similar experiments but added a finger touch this
time.
\figref{fig:rasp_touch_notouch} shows the experiment
results. Under the low voltage signal of 20 Vpp, the false touch
events occur only if a finger touches the screen. More importantly, we
can see that the positions of the false touches are centered on the
line where the true touch point is located. 
% XXX Here, we define the direction when a smartphone is at a portlait
These are desirable characteristics because usually, GUI buttons are aligned in
a row; e.g., CONNECT/CANCEL, YES/NO, or OK/CANCEL. Therefore, an
attacker can expect that a touch event will be scattered on a wrong
button, with a probability of $1/2$, with an assumption that the touch
events are uniformly scattered along a line. 
We note that screen orientation also matters. If a screen is in
portrait mode, scattered touch events along the vertical line may
not produce a touch on the targeted button. 
As we show in Section~\ref{sec:eval}, the direction of scattered touch
events differ among the different models. 
By making use of the device fingerprinting techniques, an attacker can
obtain the information about the model as well as the current screen 
orientation; these information will be used to check whether or not
Phantom touch generator is effective.

%% In Section~\ref{sec:eval}, we investigate the voltage and frequency
%% pairs that cause such misbehavior on the off-the-shelf smartphones.
%% In addition, we estimate the success rate of this attack using
%% Bluetooth pairing dialog box showed by an NFC tag.

\subsection{Electrical touch}
\label{sec:leveraging_device_design}

\begin{figure}[tbp]
    \begin{center}
        \includegraphics[width=70mm]{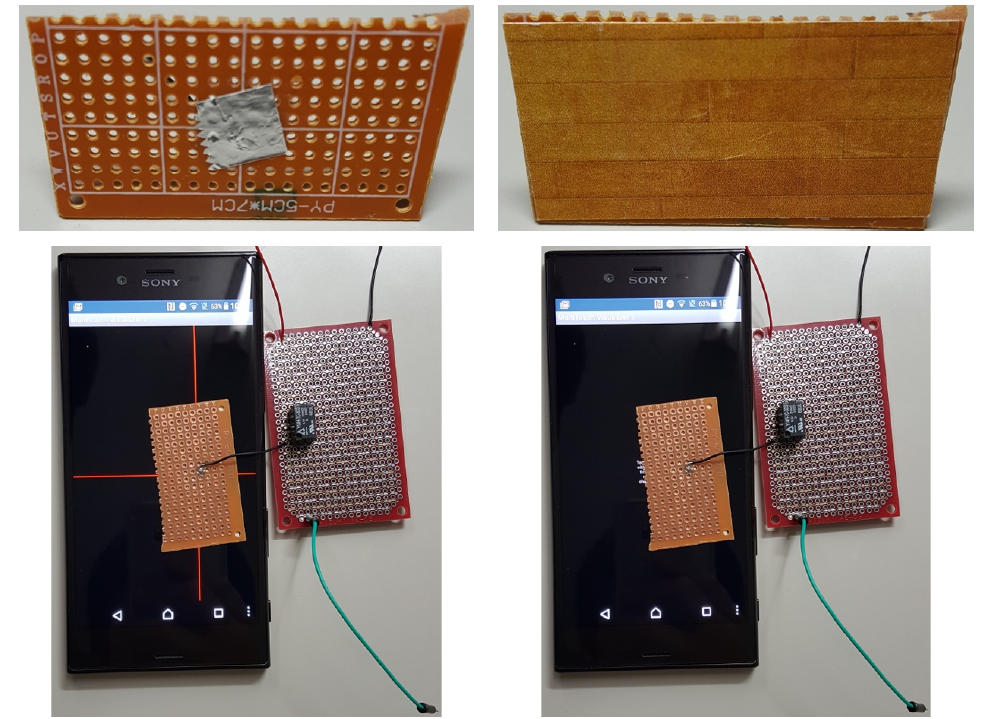}
        \caption{A circuit board that triggers the electrical
          touches. {\bf Top}:
          a plate electrode on the circuit board (left) and the
          circuit board masked with a woodgrained paper (right). {\bf
            Bottom}: A touch occurs when the plate electrode is
          connected to the green code, which has a large capacitance
          (left). A touch does not occur when the plate electrode is
          disconnected from anything (right). A simple relay can
          switch the two states: connected and disconnected.}
        \label{fig:touch_circuit}
    \end{center}
\end{figure}

So far, we have assumed that the NFC controller is mounted on the rear
side of the smartphone, and Phantom touch generator attacks the
touchscreen from the rear side. 
However, there are several smartphones/tablets that mount the NFC
controller on the front side of the devices; e.g., Nexus 10,
Xperia XZ, and ZenFone 3 Deluxe. 
These mobile devices read NFC tags on the same side with its
touchscreen. 

For this type of devices, an attacker can trigger arbitrary touch
events by directly touching the screen when a victim puts
the device on a table-type {\em ToT} with the touchscreen down.
An attacker installs a simple circuit with plate electrodes on
the surface of {\em ToT} -- embedding the circuit in the table top.
We implemented such a circuit as shown in~\figref{fig:touch_circuit}.

The circuit works as follows. 
The plate electrodes on the circuit capacitively couple with the
TX electrodes of the touchscreen when the circuit gets close enough to
the touchscreen.
The capacitance between one of the electrodes on the circuit and
TX electrodes becomes low when the circuit electrodes are
disconnected from anywhere and becomes high when the circuit electrodes
are grounded or connected to an object that has large electric capacitance.
By systemizing this mechanism, an attacker can virtually touch an
arbitrary position by relaying a corresponding electrode to the ground
or the object with a large capacitance.

The circuit shown in~\figref{fig:touch_circuit} has a
1-$\mbox{cm}^2$ of square plate electrode.
If we implement several plate electrodes that are placed 0.5 cm away
from each other, the resolution of touch becomes 1.5 cm. 
The area of plate electrode of the circuit is proportional to
the capacitance between the electrode of the circuit and
touchscreen. Therefore, plate electrodes that are too small cannot
create enough change of capacitance by grounding it. 
However, an attacker could obtain finer resolution by using circuits
used for active styluses, which actively interrupt field coupling
between the electrodes of the touchscreen.

% ソレノイドならより安定してタッチを起こせると考えられる。
% もしくはソレノイドを使用し、極板をタッチスクリーンに機械的に近づけることも考えられる。
%アクティブスタイラスの信号をリレーすると、解像度をあげられる

%% \noindent
%% \textbf{Threat model:}
An attacker can make use of this attack as follows: 
An attacker first employs the device fingerprinting to know that the
device has the NFC controller at front side.  
Using the website, the attacker can obtain the information about the
device orientation using the web API interface.
Using the position of the used NFC tag and the orientation
information, the attacker can estimate the area of the touchscreen. 
Finally, an attacker can pinpoint the position of the button for
establishing the attack and make the electrical touch by grounding the
corresponding plate electrode.

\section{Feasibility Studies}
\label{sec:eval}
To demonstrate the feasibility of ToT devices, we performed two 
empirical studies. The first study aims to verify that NFC tags
embedded inside a thing can be actually read by smartphones. 
For this study, we use 24 Android smartphones/tablets, which are
manufactured by the 12 different vendors.
We summarize the list of devices we used
in~\tabref{tab:result_studies} (Appendix).
The second study aims to verify the success of {\em Phantom touch generator}
attack.
For this study, we use 7 Android smartphones/tablets listed in
Table~\ref{tab:result_touch_flood}~\footnote{We rented 17 devices for
  the first study and were not able to used these devices for the
  second study because applying the attack has a risk of causing
  physical damages on the devices.}.

\subsection{Maximum NFC Reading Distance}
We study the maximum NFC reading distance of the smartphones
to demonstrate the validity of the idea of embedding malicious NFC
tags in a thing.
A NFC tag is attached to the backside of the wood board of the walnut
material. 
We read the tag using the smartphones placed on the backside.
We measured the maximum communicable distance by changing the
thickness of the wood board at intervals of 5 mm and recording the
success of reading the tag.
We found that the maximum NFC reading distance was 3.4 cm in
average. The maximum and minimum of the measured distance were 5.0 cm and 2.0
cm, respectively. 
The full result is summarized in~\tabref{tab:result_studies} (Appendix).
If we consider the thickness of common objects such as a table top or
a wallet, we can conclude that the measured maximum distance is large
enough to establish the attacks by ToT.

\subsection{Conditions of the successful Phantom touch generator attacks.}
\label{sec:study_touch_flood}

Using the smartphones that have the NFC controller on the frontside, 
we empirically study the conditions for the successful attacks. 
Unlike the experiment using a Raspberry Pi 7-inch touchscreen display,
we need higher voltage to establish Phantom touch generator attack. 
As our amplifier is not capable of generating voltage greater than 150
Vpp, we used a high-voltage transformer taken out of a plasma ball,
which costs about 6 USD. 

\begin{figure}[tb]
    \begin{center}
        \includegraphics[width=80mm]{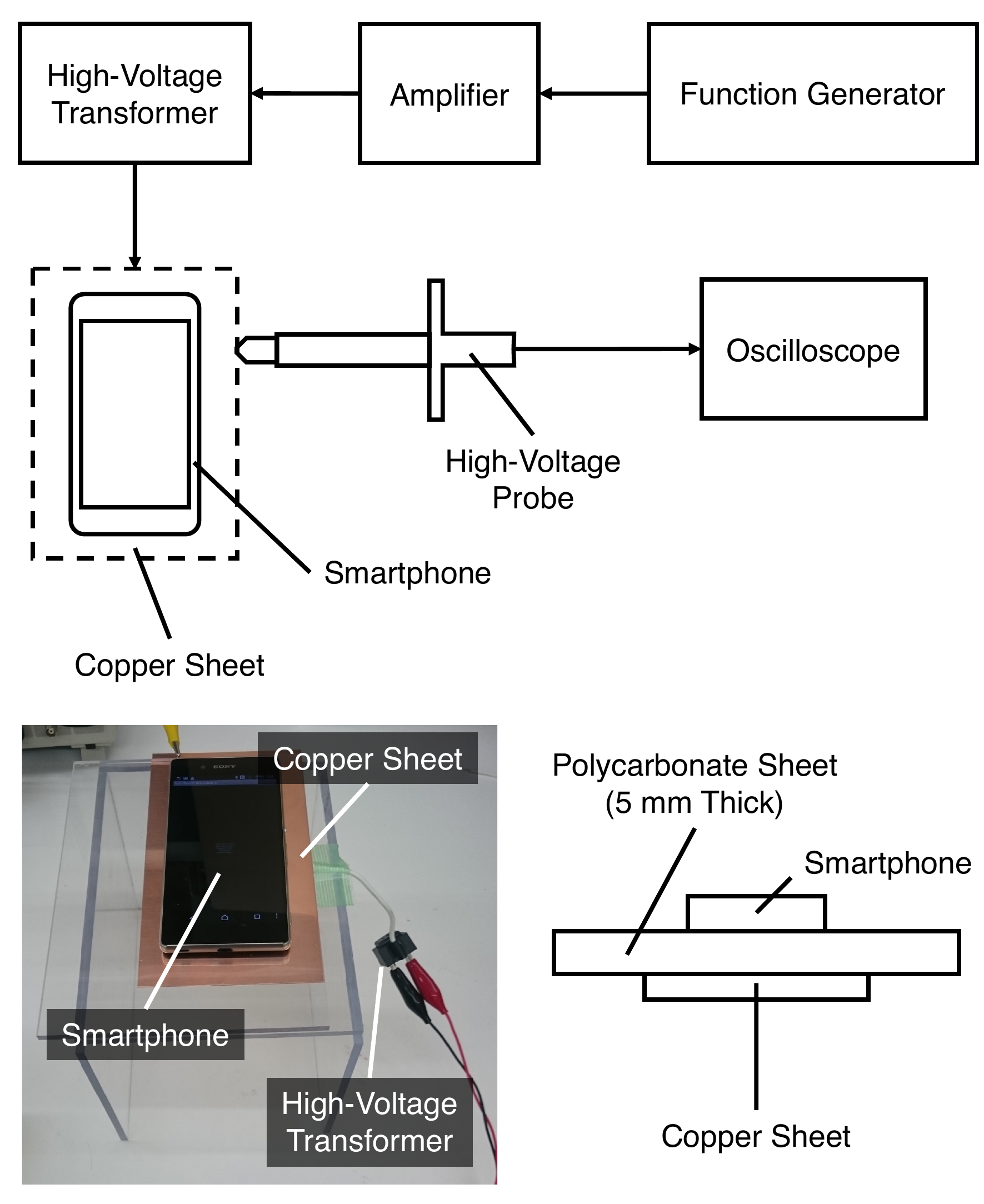}
        \caption{Block diagram and photos of setup for observing effect of alternating current on off-the-shelf smartphones}
        \label{fig:blackbox_setup}
    \end{center}
\end{figure}

\figref{fig:blackbox_setup} shows the setup of the experiments.
The smartphone and the copper sheet are insulated with the
polycarbonate plate of 5 mm thick. 
Following the procedure that is shown in Section~\ref{sec:touch_flood_attack},
we first identify the characteristic frequencies for the smartphones
to cause the malfunctions. 
For the smartphones that had caused malfunctions, we will further test
the following tasks. 
We will create a NFC tag that requests the Bluetooth paring. 
The smartphone will pop up a dialog message after reading the tag.
We then touch the button of ``NO.''
Before the smartphone reads the tag, we have applied Phantom touch
generator. 
We will see whether the actual touch becomes ``YES'' (attack succeeded)  or ``NO'' (attack failed). 

Sometimes, we do not see any responses even though we touch the
button due to the noise injection. 
In such cases, if there are no responses back after the five
consecutive touches, we count it as a failure of the attack. 
Also, if the patterns of the touch scattering for a device has a 
horizontal/vertical direction, we set the orientation of the device
to portrait/landscape. 

\begin{table*}[tb]
  \caption{Results of the {\em Touch scatterer} attack. The direction
    of the scattering patterns is defined when a screen is set in 
    portrait mode.}
  \label{tab:result_touch_flood}
  \hbox to\hsize{\hfil
    \begin{tabular}{|l|l|P{1.3cm}|R{1.7cm}|R{1.2cm}|R{1.6cm}|p{1.7cm}|}\hline
      \bf{Device} & \bf{Manufacture} & \bf{Success false touches} & \bf{Frequency \verb|[kHz]|} &
        \bf{Voltage \verb|[Vpp]|} & \bf{Success attack rates} & \bf{Scattering patterns} \\\hline\hline
      Nexus 7 & ASUS & \checkmark & 128.2 & 40.0 & 18/30 & vertical \\
      ARROWS NX F-05F & FUJITSU &  & --- & --- & --- & --- \\
      Nexus 9 & HTC & \checkmark & 280.9 & 490.0 & 0/10 & horizontal \\
      Galaxy S6 edge & SAMSUNG &  & --- & --- & --- & --- \\
      Galaxy S4 & SAMSUNG & \checkmark & 384.5 & 70.4 & 13/30 & horizontal \\
      AQUOS ZETA SH-04F & SHARP & \checkmark & 202.0 & 700.0 & 0/10 & horizontal \\
      Xperia Z4 & SONY & \checkmark & 218.0 & 340.0 & 20/30 & horizontal \\\hline
    \end{tabular}\hfil}
\end{table*}

\tabref{tab:result_touch_flood} summarizes the results. 
For the 5 out of 7 models, we specified the characteristic
frequencies and voltage values that can cause malfunction, i.e.,
``false touch.''
Of the 5 models that cause malfunctions, 3 models succeeded the attack
with probabilities distributed around $1/2$; i.e., the OS detected the
touch for a wrong button and the device was paired with a Bluetooth
device. The rest of 2 models worked as follows.
For Nexus 9, the detected touch events were biased to a specific area,
which was not close to the buttons; thus, the attacks failed. 
For AQUOS ZETA SH-04F, when a finger touched somewhere in the
right/left half of the screen, the false touches appeared on the
left/right half on the screen; thus, the attacks failed.
Thus, the patterns of false touches depend on the models.

There were two models that did not generate false touch events;
the one that the detected touch events lag behind the finger's touch
(Galaxy S 6 edge) and the one that does not recognize the touch at all
(ARROWS NX F-05F). In addition the malfunctions mentioned in
Section~\ref{sec:touch_flood_attack}, we found the following
malfunction patterns:
Even when the noise injection is stopped, the device stops reacting to
the touch until it goes to sleep mode, the monitoring application
is abnormally killed, and the operating system restarts, etc.

\section{Discussion}
\label{sec:discuss}

In this section, we discuss the feasibility of ToT attacks and
possible defenses against them.

\subsection{Feasibility}
\label{sec:feasibility}

Our threat model makes three assumptions about a victim's smartphone:
(1) It is an Android smartphone equipped with NFC, (2) the NFC
functionality is enabled on the smartphone, and (3) the screen of the
smartphone is unlocked when the smartphone is brought close to a
ToT. We now discuss the feasibility of ToT attacks in light of these
three assumptions.

The first assumption limits the scope of target devices. In fact,
although iOS supports NFC technology, it has not supported reading NFC
tags as of February 2017. Still, we conjecture that the threat of ToT
attacks is potentially pervasive in future because of the following
two reasons. First, it has been forecasted that the shipments of
Android NFC-enabled smartphones will reach 844 million in 2018~\cite{nfcphoneforecast},
indicating that the potential target of ToT attacks is increasingly
becoming ubiquitous. 
In addition, there is a possibility that Android Instant
Apps~\cite{instant-apps} accelerates the adoption of NFC. 
We note, however, that the naive use of the new technology has
potential security risks such as launching a fake browser, etc.
Second, many financial technology companies have
recently launched mobile payment services using NFC technology; this
trend will continue to grow and push the adoption of NFC-empowered
smartphone services.

The second assumption limits the opportunities of successful attacks;
i.e., the attack will not succeed unless the NFC functionality is
enabled on the victim's smartphone. To verify the second assumption,
we manually investigated 24 smartphones listed in
Table~\ref{tab:result_studies} (Appendix).
We found that the NFC functionality is enabled in the
factory setting in 16 out of 24 models. Interestingly, in more recent
models, the NFC functionality is enabled in the factory setting. The
results of our survey are summarized in the appendix (see
Table~\ref{tab:result_studies}). As we already have discussed before,
we also conjecture that the number of users who enjoy NFC services on
their smartphone (thus, will enable NFC) will keep on increasing.

Finally, the third assumption also limits the opportunities of
successful attacks; i.e., even if a smartphone approaches a ToT, the
attack will fail if the smartphone's screen is locked: Android OS
will not invoke functionalities recorded in the NDEF record when the
screen is locked. To verify the attack feasibility, we analyzed two
types of ToT, a simple ToT and {\em ToT device}. Since a simple ToT
can be disseminated as a small thing with an NFC tag attached, the
attacker can easily produce a large number of ToTs with a reasonable
cost. Thus, a simple ToT has high affinity with mass attack; producing
more ToTs will increase the expected number of successes even
if the probability of each attack is small.

In contrast, it will not be easy for an attacker to install {\em ToT
  device} in many places due to the cost issues. The key success
factor of {\em ToT devices} is attributed to the patterns of human
behavior. Many people use smartphones while eating food or drinking
coffee. If the attacker installs a table-type {\em ToT device} in an
eatery or a coffeehouse, the probability of the {\em ToT device}
encountering a smartphone placed on the table with the screen unlocked
is high. There are other situations when a person places a smartphone
on a desk or a table without locking the screen. If such a table is
installed at a public space such as library, many people will use the
table in a day. Of these, there will be several who own NFC-enabled
Android phones and place it on the table without unlocking the screen,
i.e., the expectation value of the number of successful attacks
becomes high. The malicious table will keep waiting for new victims as
long as power is supplied. In our future work, we plan to conduct
field studies to quantify the correlation between human behavior
patterns and the success of an attack.

%% \textbf{Things:}
%% %TODO: 合ってる？
%% %NFCは電磁誘導を利用した通信を行うため，磁気シールド効果のある物体中に埋め込むことはできない．
%% %生じた渦電流により磁束が打ち消される？
%% 悪性NFCタグによる攻撃の機会を多く得るには，
%% タグが埋め込まれていることを気づかれることなく，
%% モノが使用される必要がある．
%% 特に貨幣型トロイを実装するためには実際に流通している通貨に悪性NFCタグを埋め込む必要があるが，
%% 手触りや厚みが変化することは避けられず，
%% 通貨使用者が加工に気づきやすいと考えられる．
%% このように加工したことが気づかれやすいモノに，悪性NFCタグを埋め込むことは合理的でない．
%% しかし，通貨の加工に気づいても，
%% その加工が悪意のあるものであると判断するとは限らない．
%% 例えば，アメリカでは紙幣にシールを貼ることが許されており~\cite{celebrity_notes}，
%% 実際にサンタクロース等のシールが貼付されたドル紙幣が販売されている~\cite{santaDollars}．
%% 通貨へのシール貼付が許されており，それが一般に認知されている場合，
%% 貨幣型トロイが長期間流通してしまう懸念がある．

%% \noindent
%% \textbf{For Your Safety:}

%% \noindent
%% \textbf{Etc:}
%% Android Instant Appsの実装によっては、
%% ScreenFilter等のアプリがあらかじめインストール済みでなくても攻撃可能

%% 攻撃者が端末の物理的位置を仮定できるのであれば、
%% タッチスクリーンを用いたユーザーへの確認は効力が弱くなる。

\subsection{Countermeasures}
We now discuss possible countermeasures against the threat of ToT
attacks. We divide the discussion into three groups according to three
points of view.

\paragraph{mobile OS:}
The simplest and the most effective defense is to add/improve the user
approval processes before the mobile OS launches applications recorded
in a tag. For instance, by forcing to request user approval for all
NFC-driven operations shown in Table~\ref{tab:nfc-android}, we can eliminate the threats
of a simple ToT that leverages the single-shot attack, which targets
operations that do not require approval. Even when the attacker uses
a {\em ToT device}, showing a proper message will decrease the chances
of attack success. To this end, mobile OS vendors should change the
format of messages associated with NDEF records.
By explicitly presenting the reason why an operation is invoked, it is
possible to create a message format that makes it impossible to
generate deceptive messages.

Making the user approval process more rigorous could sacrifice the
usability of NFC-powered services. To solve this problem, we can
leverage the context of NFC touch events. This has been explored by
Czeskis et al.~\cite{czeskis2008rfids}, who developed techniques to achieve context-aware
communication for RFID tags and contact-less cards. Their key idea was
to leverage the built-in accelerometer, which can be used to implement
activity recognition techniques to infer whether or not the holder of
the tag physically moves her/his hands, e.g., tap the tag against the
reader. We can use a similar technique to distinguish legitimate touch
events from the false events generated by a ToT. 
Smartphones have other sensors that can be used to infer
the context of touch events, e.g., proximity sensor and illuminance
sensor. If the smartphone infers that the context is likely an attack,
the level of user approval can be increased to give priority to
security; otherwise, the level of user approval is decreased to give
priority to usability. Likewise, if a smartphone infers from sensor
data that it is likely to be inside a pocket, it can automatically
lock the screen to prevent the device from reading NFC tags
unintentionally.

\paragraph{Smartphone hardware:}

While conducting the experiments described Section~\ref{sec:eval}, we
noted that
some touchscreen controllers stopped working when a strong electric
field was applied. Although these observations are not conclusive, we
conjecture that the manufactures of these controllers may have
installed mechanisms to stop the controllers upon detection of
external noises. In fact, as Ref.~\cite{cypress_noise_immunity}
reported, manufactures of
touchscreen controllers have developed techniques for dealing with the
noise that can interfere with capacitive touch sensing. Incorporating
such mechanisms will lead to eliminating the threats of touch
scatterer. 
In addition, as Kune et al. proposed in Ref.~\cite{kune2013ghost},
there are several analog/digital countermeasures against intentional
EMI attacks, e.g., a filter that attenuates external noise signals and
signal processing to eliminate anomalous inputs. These techniques will
also be useful as countermeasures against the threats of Phantom touch
generator.
In Section~\ref{sec:last_touch}, we also demonstrated that design of
mounting the NFC controller on the front side of the smartphone makes
generating false touch attacks easy. To defend against the threat of
such attacks, mounting the NFC controller on the back is more
desirable.

\paragraph{Things:}
It is almost impossible to visually detect a ToT because NFC tags are
embedded into physical things. However, there may be situations where
law enforcement agencies want to inspect tables inside a building to
investigate whether a ToT has been installed. An active probe that
searches for NFC tags should be developed to make this task
easier. For this purpose, it is also possible to build a ToT honeypot
that behaves as an NFC-enabled smartphone. The drawback of this
approach is that it is not scalable because the practical working
distance range of NFC is at most about 4 cm.
Further research is needed to shed more light on this problem.

%% \begin{figure}[tb]
%%     \setbox0\vbox{
%%         \hbox{$\cdots$}
%%         \hbox{\footnotesize \verb|<string name="prompt_connect_to_network" msgid="8511683573657516114">|}
%%         \hbox{\footnotesize \verb|"NFCタグからWi-Fi設定データを受信しました。\n|}
%%         \hbox{\footnotesize \verb|以下のSSIDのWi-Fiネットワークに接続しますか？\n|}
%%         \hbox{\footnotesize \verb|SSID:[<xliff:g id="NETWORK_SSID">%1$s</xliff:g>]"|}
%%         \hbox{\footnotesize \verb|</string>|}
%%         \hbox{$\cdots$}
%%     }
%%     \centerline{\fbox{\box0}}
%%     \caption{改良した Nfc/res/values-ja/strings.xml}
%%     %\ecaption{improved Nfc/res/values-ja/strings.xml.}
%%     \label{fig:counter_ssid_string}
%% \end{figure}

%% \subsection{Ethical considerations}
%% As several researchers have reported, the threat of attacks using
%% malicious NFC tags is publicly known~\cite{miller2012don,mulliner2009vulnerability,wallofsheep,gold2015testbed,verdult2011practical,rieback2006your}.
%% Several groups such as the NFC Security Awareness Project and W3C have
%% addressed the danger of reading unknown NFC
%% tags~\cite{wallofsheep,w3c-nfc}. The objective of our work was to
%% explore the threats of malicious NFC tags by embedding them into
%% common objects. As we have demonstrated, the threats of malicious NFC
%% tags become further viable through a ToT. Although our attack is a
%% proof-of-concept, we provide possible countermeasures that will remedy
%% the threats. We hope that our paper will be a catalyst to further
%% enhance the security of NFC-powered smartphones.

\section{Related Work}
\label{sec:related}

\noindent\textbf{Attacks using NFC tags:}
There have been several studies on the threats of attacks using NFC
technology~\cite{rieback2006your,mulliner2009vulnerability,verdult2011practical,miller2012don,wallofsheep,gold2015testbed}. 
%verdult2011practical(Bluetooth Pairing)とmulliner2009vulnerability(NFC worm)はNokia 6000Seriesが対象、rieback2006your(SQL injection)はPCが対象なので説明を省く
Miller~\cite{miller2012don} reported that malicious NFC tags can
attack browser exploits and NFC stack bugs that existed at the time.
Gold et al.~\cite{gold2015testbed} demonstrated a phishing attack that
uses a smart poster with malicious NFC tag attached. The accessed
website prompts users to log in to a fake SNS site.
They also demonstrated that an attacker can write a malicious file to
the victim's device by using the peer-to-peer mode of NFC.
Wall of Sheep~\cite{wallofsheep} demonstrated the experiment using
NFC tags attached to smart posters and buttons at the DEFCON venue. 
At the venue, they put posters that say ``Find a Wall of Sheep button
and scan it with your NFC phone for exclusive discounts, tools and
surprises every day.'' They reported that about 50 attendees scanned
the NFC tags that ``could'' have been malicious tags.
These studies assumed that an attacker can come close enough to
a victim, or the victim intentionally read the malicious NFC tag by
using posters or other existing facilities.
The threat model is different from the one for ToT; an attacker
injects malicious NFC tags into common objects.  

\noindent\textbf{RFID tags:}
NFC is a specialized subset within the family of radio frequency
identification (RFID) technology.
Several researchers have studied the risk of RFID tags that can be
attached to various things~\cite{baldini2005rfid,juels2006rfidsec}. 
Baldini et al.~\cite{baldini2005rfid} reported the application of RFID
tags in the retail sector and discussed associated privacy issues and
countermeasures.
Juels published a survey paper on the research of privacy and security
of RFID~\cite{juels2006rfidsec}. The survey examined the privacy
protection mechanisms and integrity assurance in RFID systems. 
In the paper, Juels mentioned the importance of {\em user perception}
of security and privacy in RFID systems as users cannot see RF
emissions. The indication is closely related the problem we addressed
in this paper. 

The absence of user perception in RFID systems leads to the ``relay
attack,'' which enables an attacker to sets up a link between the
reader and the contactless card without the agreement of the owner.
Several countermeasures against the relay attack on RFID systems have
been studied~\cite{heydt2007vulnerabilities,czeskis2008rfids,francillon2011relay,mehrnezhad2015tap}.
As we discussed in Section~\ref{sec:discuss}, the techniques used as
the countermeasures against relay attack, such as context-aware
communication, can be useful to tackle the threats of ToT.

%%pc: srdjan capkun
%%本研究のものはPhishingなのか？
%スマートフォンに対するフィッシング攻撃には様々なものがあるが\cite{Marforio:2016:EPS:2858036}、
%本研究で提案したフィッシング攻撃(?)は(Confirmation message trap, Exploiting Installed apps)、
%良性のアプリやOSのダイアログを利用したものであり、既存の対策手法では対抗できないものである。

\noindent\textbf{Attacks on touchscreen:}
There have been many studies on the side-channel attacks on
touchscreens (LCDs); Aviv et al.~\cite{aviv2010smudge} used smudge
left on the screen to infer a graphical password, Maggi et
al.~\cite{maggi2011fast} used the data collected from a surveillance
camera to recognize keystrokes of a victim, and Hayashi et
al.~\cite{Hayashi:2014:TTP:2660267.2660292} used electromagnetic
emanation to reconstruct a victim's tablet display.
To the best of our knowledge, while these attacks passively steal
data from the touchscreen, our {\em Phantom touch generator} is the
first attack that actively radiates signals toward touchscreen to
cause targeted malfunctions.  

% PCのKevin Fuさんの論文であり、 
% EMIへの(アナログ、デジタル両面での)対策手法がたくさん載っているので
% 引用できると良いと思います ~\cite{kune2013ghost}

\section{Conclusion}
\label{sec:conclusion}
We introduced a novel proof-of-concept attack named {\em ToT}, which
targets NFC-enabled smartphones.
The key concept of ToT is to inject malicious functionalities
into common objects, which are {\em not} considered as NFC touchpoints.
We believe that this concept sheds new light on the security research
of mobile/IoT devices.
To fully explore the threats of ToT attacks, we developed two
effective techniques: 
{\em ToT device} and
{\em Phantom touch generator}, which enable an attacker to carry out
various
severe and sophisticated attacks without being perceived by the device
owner who unintentionally puts the device close to a ToT.
Through the extensive experiments using off-the-shelf smartphones, we
demonstrated that the proposed attacks work in practice.
Although our attack is a proof-of-concept, we provide possible
countermeasures that will thwart the threats. We hope that our paper
will be a catalyst to further enhance the security of NFC-powered
smartphones.

\section*{Acknowledgements}
We thank Prof. Yasushi Matsunaga, Prof. Yuichi Hayashi, Prof. Masahiro
Kinugawa, Prof. Yoshimichi Ohki, and Mr. Kazuyuki Ishimoto
for sharing the valuable comments on the mechanism of Phantom touch
generator. 
We also thank Mr. Makoto Gotoh for granting permission to use the
several equipments we used for the experiments of Phantom touch
generator.
We are greatful to Prof. N. Asokan for providing us with useful
comments and feedback to the earlier versions of this manuscript.
Finally, we thank Prof. Shigeki Goto, Mr. Yumehisa Haga, and
Mr. Haruka Hoshino for their various cooperation in carrying out this
work.

\bibliographystyle{acm}
\bibliography{bibliography}

\newpage
\newpage
\appendix

\section{Supplemental data.}
In this section, we provide supplemental data that are ommited
in the main body of the paper due to the space limitation.

%\subsection{Screenshots of Wi-Fi connection dialog box.}

\begin{figure}[htbp]
  \centering
  \includegraphics[width=60mm]{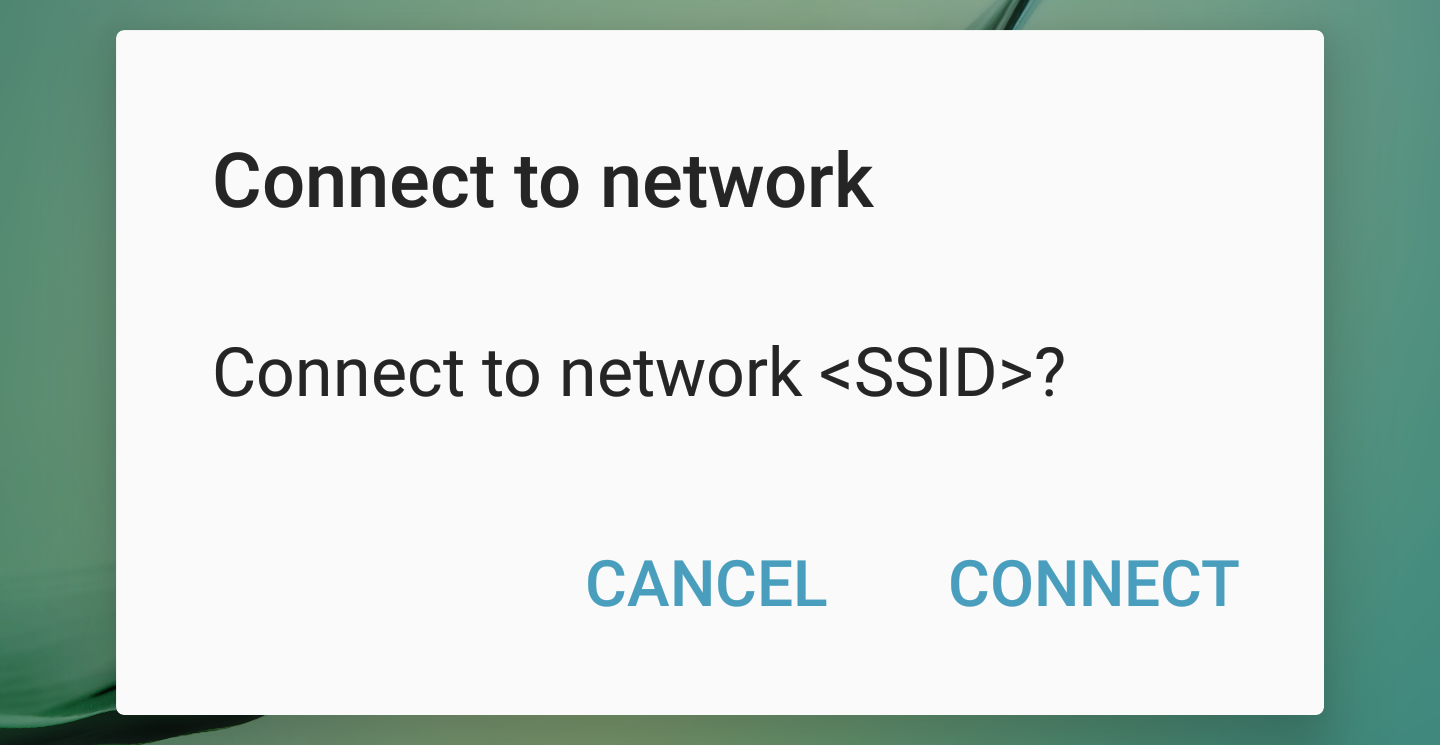}
  \caption{Wi-Fi connection dialog box (normal).}
  \label{fig:ssid_normal}
\vspace{5mm}
  \centering
  \includegraphics[width=60mm]{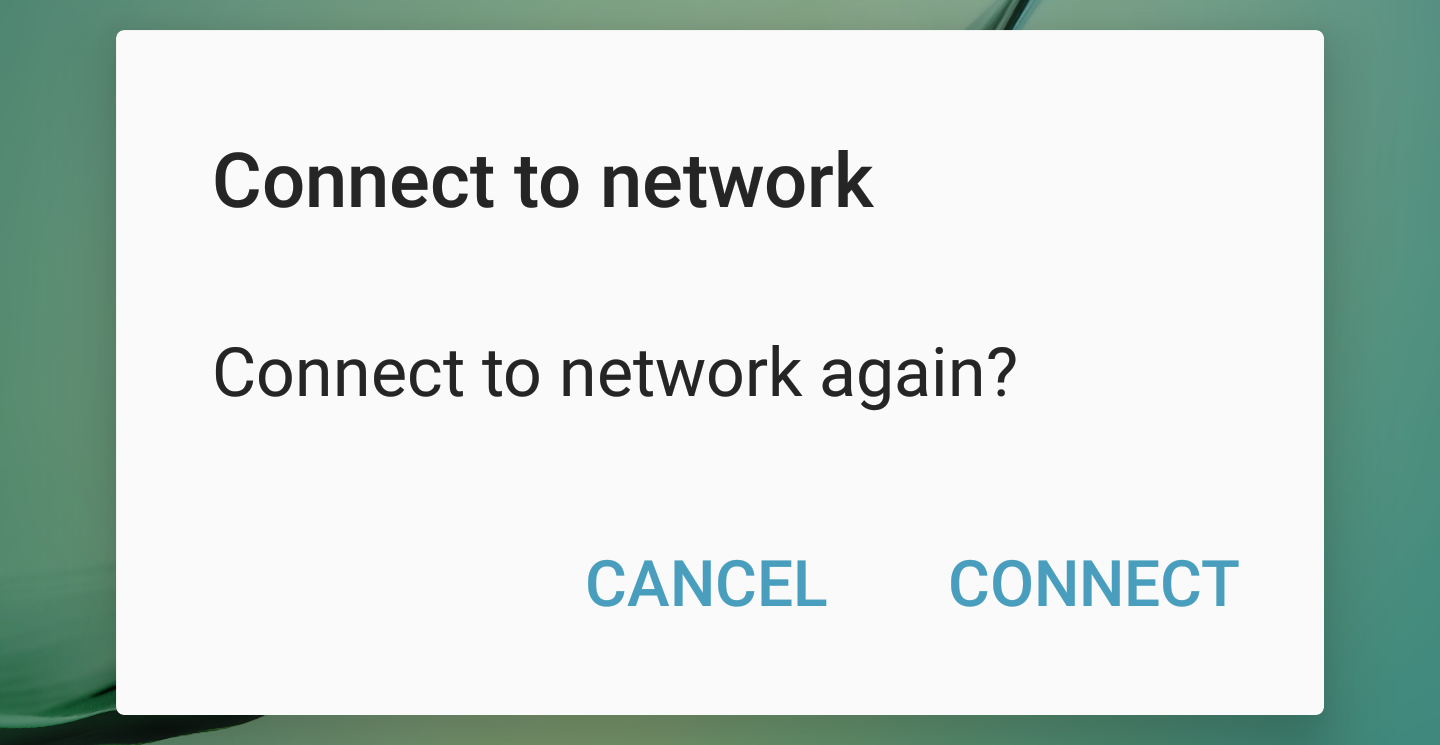}
  \caption{Wi-Fi connection dialog box (attacked).}
  \label{fig:ssid_spoof}
\vspace{5mm}
  \centering
  \includegraphics[width=60mm]{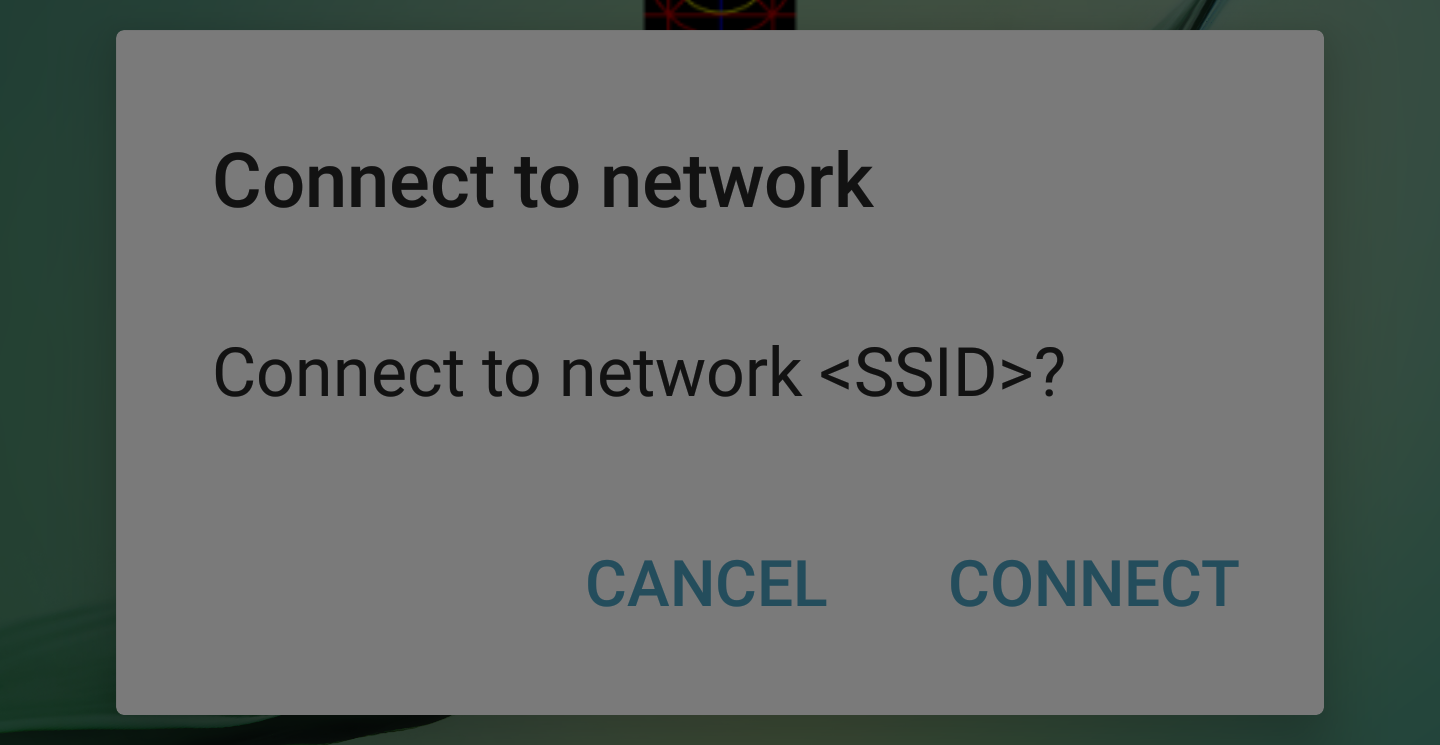}
  \caption{Wi-Fi connection dialog box (dimmed using Screen Filter app).}
  \label{fig:ssid_screenfilter}
\vspace{5mm}
  \centering
  \includegraphics[width=60mm]{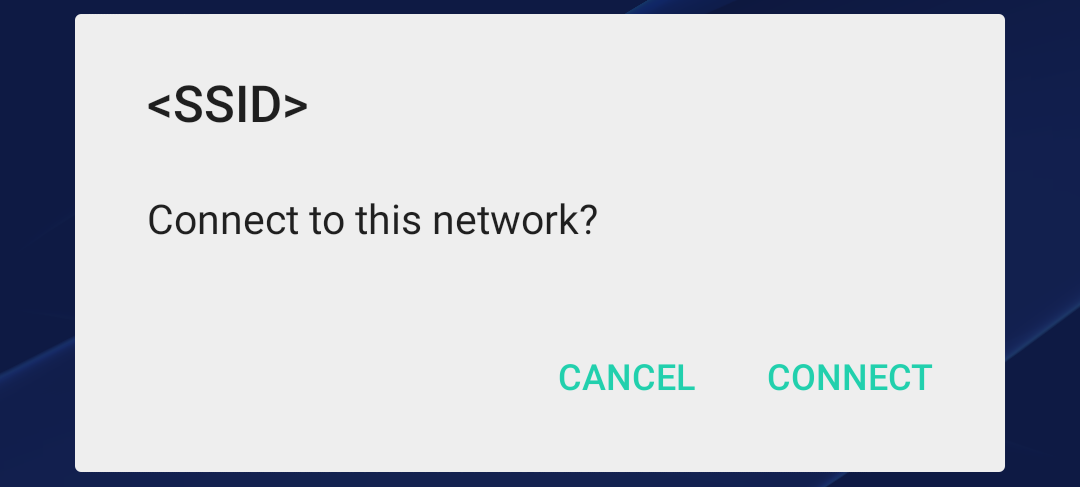}
  \caption{Wi-Fi connection dialog box (customized for Xperia Z3).}
  \label{fig:ssid_xperiaz3}
\end{figure}

\begin{figure}[htbp]
  \centering
  \includegraphics[width=55mm]{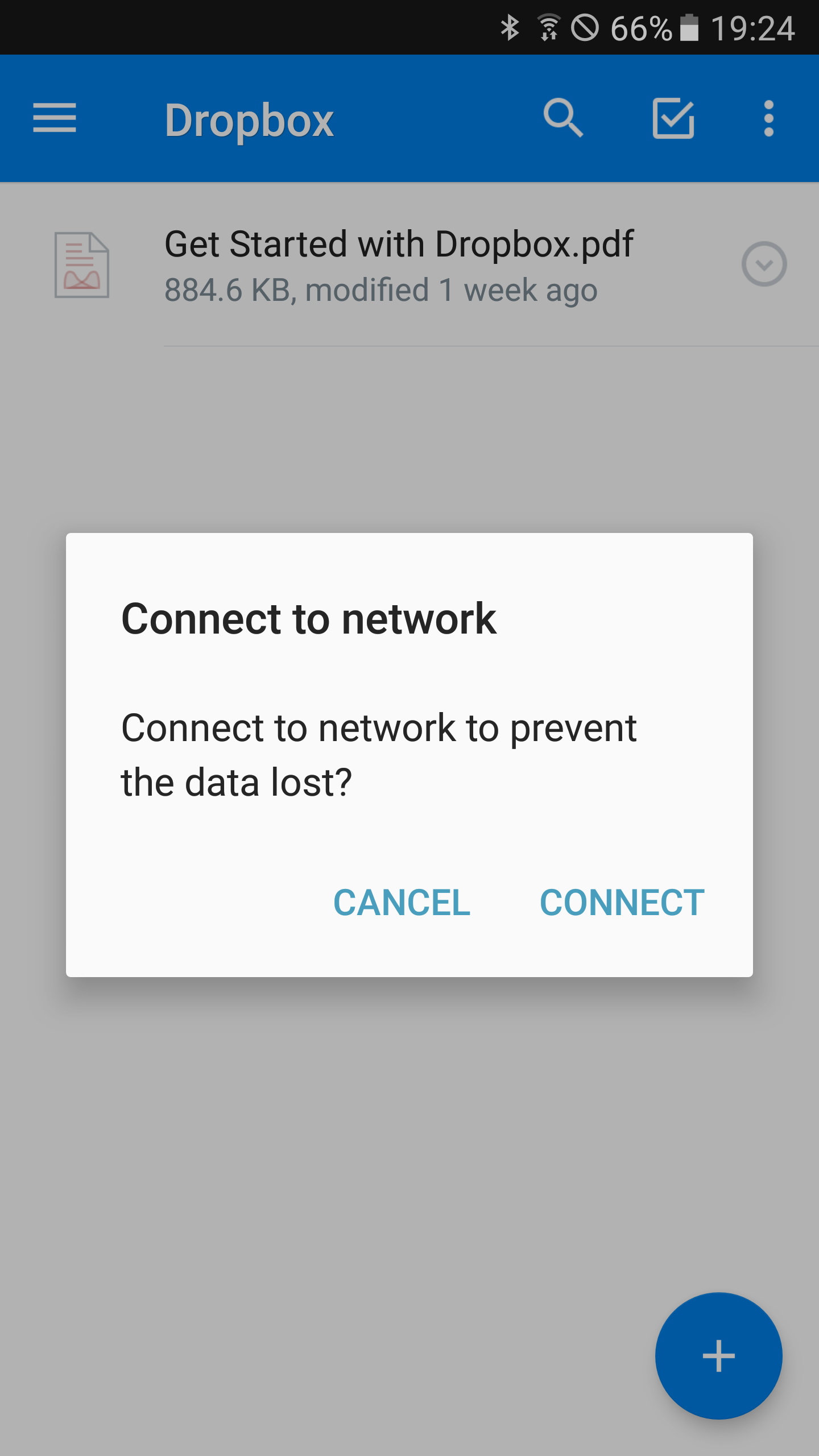}
  \caption{Wi-Fi connection dialog box (attack using Dropbox app).}
  \label{fig:ssid_dropbox}
\vspace{5mm}
  \centering
  \includegraphics[width=55mm]{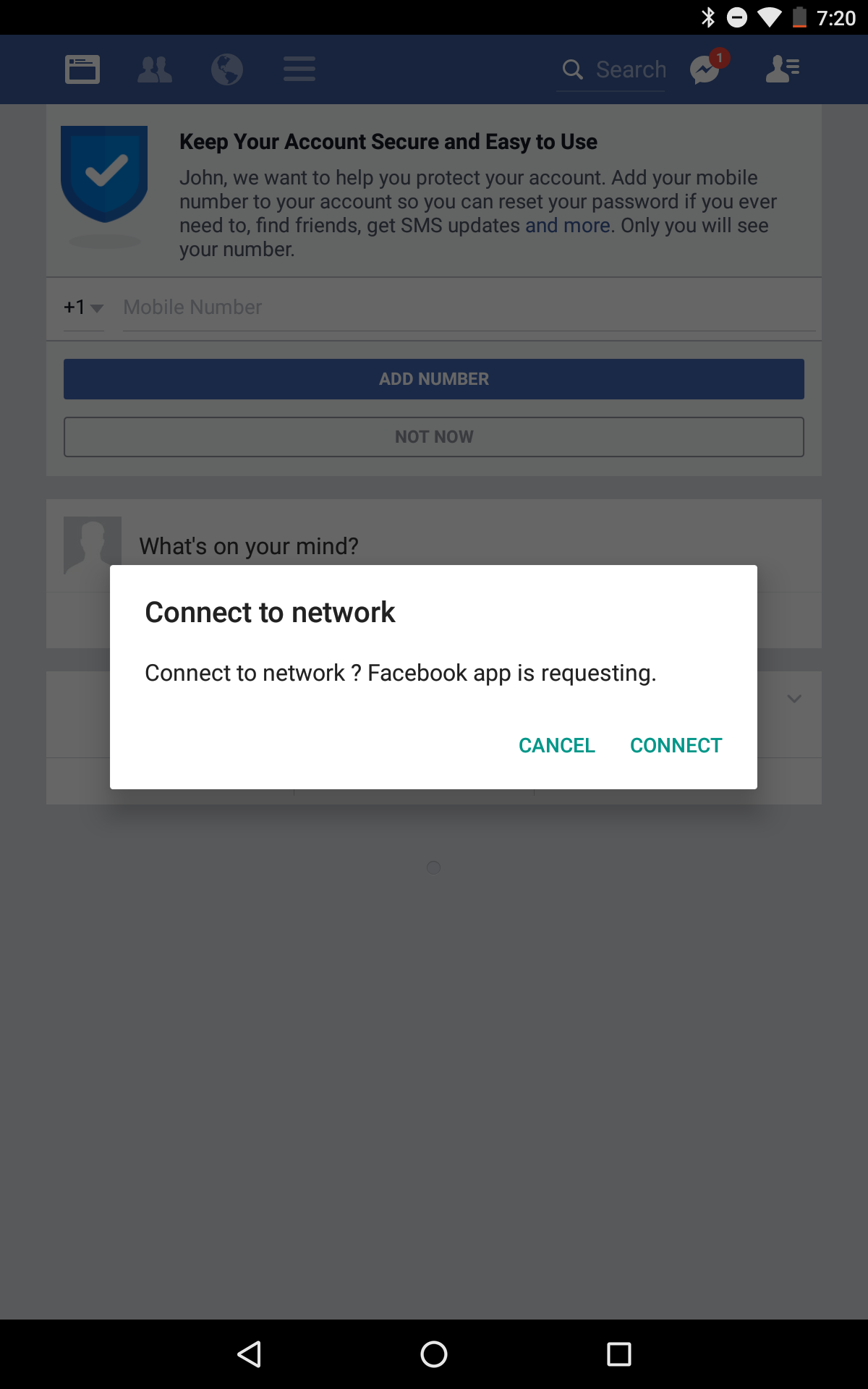}
  \caption{Wi-Fi connection dialog box (attack using Facebook app).}
  \label{fig:ssid_facebook}
\end{figure}

%\subsection{Result of the feasibility test.}

%% Using the devices listed in Table~\ref{tab:smartphone}, 
%% we investigated the 

%% 工場出荷時のNFC設定，およびNDEFレコードにより表示される確認ダイアログ
%% NDEFレコードにより表示される確認ダイアログの表示内容については，\tabref{tab:result_wifi}，\tabref{tab:result_bt}に示す．
%% \tabref{tab:result_wifi}ではWiFiConfigレコードに設定したSSIDを``\verb|<SSID>|''で，
%% \tabref{tab:result_bt}ではBTSSPレコードに設定した機器の名称を``\verb|<name>|''でそれぞれ示した．
%% WiFiConfigレコードはAndroid 5.0以降よりサポートされている．そのため，Androidのバージョンが5.0未満の端末では
%% WiFiConfigレコードにより表示されるメッセージは存在しない（\tabref{tab:result_smartphones}，\tabref{tab:result_range_message}）．

%% また工場出荷状態において，過半数のAndroid端末でNFCのリーダ/ライタモードが有効に設定されており，
%% 悪性NFCタグによる攻撃が可能な状態となっていた．
%% さらに，確認ダイアログのメッセージは，``Ascend P7''を除いて，NFCタグにより確認ダイアログが表示されたのだとユーザが判断できるような内容ではなかった．
%% 特に，WiFiConfigレコードにより表示される確認ダイアログで，
%% \ref{sec:spoof_message}項の手法に対抗できるものは存在しなかった．
%% 以上より，悪性NFCタグで実装したTrojan of Thingsの脅威がきわめて現実的であると結論づけることができる．

\begin{table*}[tbp]
  \caption{Results of Feasibility Studies}
  \small
    \label{tab:result_studies}
    \hbox to\hsize{\hfil
    \begin{tabular}{|l|l|p{1.3cm}|R{1.6cm}|P{1.8cm}|p{2.2cm}|p{2.2cm}|}\hline
        \bf{Device} & \bf{Manufacture} & \bf{Android Version} & \bf{Maximum Reading Distance \verb|[cm]|} & \bf{NFC R/W Activated in Factory State} & \bf{Message~Type (Wi-Fi)} & \bf{Message~Type (Bluetooth)}\\\hline\hline
        ONETOUCH IDOL 2 S & ALCATEL & 4.3 & 3.0  &  & --- & BT-EN-1\\
        Nexus 7 & ASUS & 6.0.1 & 4.0  & \checkmark & WI-EN-1 & BT-EN-1\\
        SAMURAI KIWAMI & FREETEL & 5.1 & 3.0  &  & WI-EN-1 & BT-EN-1\\
        ARROWS NX F-05F & FUJITSU & 5.0.2 & 4.0  &  & WI-EN-1 & BT-EN-1\\
        Nexus 9 & HTC & 7.0 & 4.5  & \checkmark & WI-EN-1 & BT-EN-1\\
        INFOBAR A02 & HTC & 4.1.1 & 2.5  &  & --- & BT-EN-1\\
        Ascend P7 & HUAWEI & 4.4.2 & 3.5  & \checkmark & --- & BT-EN-4\\
        TORQUE G02 & KYOCERA & 5.1 & 3.5  & \checkmark & WI-EN-1 & BT-EN-1\\
        TORQUE G01 & KYOCERA & 4.4.2 & 3.5  & \checkmark & --- & BT-EN-1\\
        Nexus 5X & LG & 6.0 & 4.5  & \checkmark & WI-EN-1 & BT-EN-1\\
        isai vivid & LG & 5.1 & 5.0  & \checkmark & WI-EN-2 & BT-EN-2\\
        DM-01G & LG & 5.0.2 & 5.0  &  & WI-EN-2 & BT-EN-2\\
        ELUGA P & PANASONIC & 4.2.2 & 2.0  &  & --- & BT-EN-1\\
        Galaxy S7 edge & SAMSUNG & 6.0.1 & 3.0 & \checkmark & WI-EN-1 & BT-EN-5\\
        Galaxy S6 edge & SAMSUNG & 6.0.1 & 2.0 & \checkmark & WI-EN-1 & BT-EN-5\\
        Galaxy S4 & SAMSUNG & 5.0.1 & 3.0  &  & WI-EN-1 & BT-EN-5\\
        AQUOS ZETA SH-01H & SHARP & 5.1.1 & 3.5  & \checkmark & WI-EN-1 & BT-EN-1\\
        AQUOS ZETA SH-04F & SHARP & 5.0.2 & 3.5  & \checkmark & WI-EN-1 & BT-EN-1\\
        AQUOS SERIE & SHARP & 5.0.2 & 3.0  & \checkmark & WI-EN-1 & BT-EN-1\\
        Xperia XZ & SONY & 7.0 & 3.0  & \checkmark & WI-EN-1 & BT-EN-3\\
        Xperia Z5 & SONY & 6.0 & 3.0  & \checkmark & WI-EN-1 & BT-EN-3\\
        Xperia Z4 & SONY & 6.0 & 4.0  & \checkmark & WI-EN-1 & BT-EN-3\\
        Xperia Z3 & SONY & 5.0.2 & 3.0  & \checkmark & WI-EN-3 & BT-EN-3\\
        Xperia Z2 & SONY & 5.0.2 & 2.5  &  & WI-EN-3 & BT-EN-3\\\hline
    \end{tabular}\hfil}
\end{table*}

\begin{table*}[htbp]
    \caption{List of confirmation messages invoked by the WiFiConfig record}
    \label{tab:messages_wifi}
    \hbox to\hsize{\hfil
    \begin{tabular}{|l|l|l|l|l|}\hline
        \bf{Type} & \bf{Title} & \bf{Message} & \bf{Positive Button} & \bf{Negative Button}\\\hline\hline
        WI-EN-1 & Connect to network & Connect to network \verb|<SSID>|? & CONNECT & CANCEL\\
        WI-EN-2 & Connect & Connect to \verb|<SSID>|? & YES & NO\\
        WI-EN-3 & \verb|<SSID>| & Connct to this network? & CONNECT & CANCEL\\\hline
    \end{tabular}\hfil}

    \caption{List of confirmation messages invoked by the BTSSP record}
    \label{tab:messages_bt}
    \hbox to\hsize{\hfil
    \begin{tabular}{|l|l|l|p{1.4cm}|p{1.4cm}|}\hline
        \bf{Type} & \bf{Title} & \bf{Message} & \bf{Positive Button} & \bf{Negative Button}\\\hline\hline
        BT-EN-1 & --- & Are you sure you want to pair the Bluetooth device ? & YES & NO\\
        BT-EN-2 & --- & Bluetooth pairing requested. Pair? & YES & NO\\
        BT-EN-3 & --- & Pair with [\verb|<name>|]? & YES & NO\\
        BT-EN-4 & NFC pairing request & Pair with the Bluetooth device ? & Pair & Cancel\\
        BT-EN-5 & --- & Pair the Bluetooth device ? & YES & NO\\\hline
    \end{tabular}\hfil}
 \end{table*}

%% \section{NFC Detection Area}
%% % Xperia XZに行うべきでは？
%% \tabref{tab:result_range_message}で示したNFC最大通信可能距離を測定するために行った実験の結果の一例として，
%% Xperia~Z3についてNFC通信可能領域を測定した結果を図示したものを\figref{fig:read_range}に示す．

%% 測定では，NFCタグの貼付位置を1 cm間隔で動かしながら，貼付位置ごとに3回の試行を行った．
%% \figref{fig:read_range}の黒い領域は，その領域内にNFCタグの中心点が位置しているときに，
%% NFCタグがスマートフォンに1回以上読み込まれたことを示している．
%% また，星のマークはXperia~Z3に刻印された``モバイル非接触IC通信マーク''の位置を示している．
%% モバイル非接触IC通信マークは，スマートフォンがNFCタグを読み込む位置の目印としてメーカーにより刻印されているものである．

%% \begin{figure}[t]
%%     \begin{center}
%%         \fbox{
%%             \includegraphics[width=75mm]{figs/read_range.pdf}
%%         }
%%         \caption{スマートフォンとNFCタグの通信可能領域の測定結果（Xperia~Z3）}
%%         %\ecaption{Communication range between a smartphone and a nfc tag.}
%%         \label{fig:read_range}
%%     \end{center}
%% \end{figure}

%% \section{Japanese Message Definition of NFC Dialog Box}

%% "ARROWS NX"->"ARROWS NX F-05F"

%\theendnotes

\end{document}